\documentclass[12pt,draftcls,onecolumn]{IEEEtran}

\usepackage{mathrsfs}

\usepackage{graphicx}
\usepackage{color}
\usepackage{amsmath,amsfonts,amssymb}
\usepackage{cite}
\usepackage{alltt}
\usepackage{algorithm,algorithmic}
\usepackage{epstopdf}
\usepackage{multirow}
\usepackage{natbib}
\usepackage{url}

\begin{document}

\newtheorem{theorem}{Theorem}
\newtheorem{lemma}{Lemma}
\newtheorem{definition}{Definition}
\newtheorem{remark}{Remark}
\newtheorem{corollary}{Corollary}
\newtheorem{example}{Example}
\newtheorem{assumption}{Assumption}

\title{Online Algorithms for Recovery of Low-Rank Parameter Matrix in Non-stationary Stochastic Systems}
\author{Yanxin~ Fu$^{*}$,
Junbao~Zhou$^{**}$, Yu~Hu$^{**}$,
and Wenxiao~Zhao$^{*}$
\thanks{The research of Yanxin Fu and Wenxiao Zhao was supported by National Key R\&D Program of China under Grant 2024YFC3307201, National Natural Science Foundation of China under Grant 12288201, and CAS Project for Young Scientists in Basic Research under Grant YSBR-008.}
\thanks{$^*$ State Key Laboratory of Mathematical Sciences (SKLMS), Academy of Mathematics and Systems Science, Chinese Academy of Sciences, China; School of Mathematical Science, University of Chinese Academy of Sciences, China (e-mail: fuyanxin@amss.ac.cn; wxzhao@amss.ac.cn). }
\thanks{$^{**}$ Research Center for Intelligent Computing Systems, Institute of Computing Technology, Chinese Academy of Sciences, China; School of Computer Science and Technology, University of Chinese Academy of Sciences, China (e-mail: zhoujunbao22@mails.ucas.ac.cn; huyu@ict.ac.cn).}
}

\date{}
\maketitle

\vspace{-10mm}

\begin{abstract}
This paper presents a two-stage online algorithm for recovery of low-rank parameter matrix in non-stationary stochastic systems. The first stage applies the recursive least squares (RLS) estimator combined with its singular value decomposition to estimate the unknown parameter matrix within the system, leveraging RLS for adaptability and SVD to reveal low-rank structure. The second stage introduces a weighted nuclear norm regularization criterion function, where adaptive weights derived from the first-stage enhance low-rank constraints. The regularization criterion admits an explicit and online computable solution, enabling efficient online updates when new data arrive without reprocessing historical data. Under the non-stationary and the non-persistent excitation conditions on the systems, the algorithm provably achieves: (i) the true rank of the unknown parameter matrix can be identified with a finite number of observations, (ii) the values of the matrix components can be consistently estimated as the number of observations increases, and (iii) the asymptotical normality of the algorithm is established as well. Such properties are termed oracle properties in the literature. Numerical simulations validate performance of the algorithm in estimation accuracy.
\end{abstract}


\section{Introduction}
\label{sec:introduction}

\IEEEPARstart{H}{igh-dimensional} stochastic systems find extensive applications in signal processing, statistics, neural networks, and systems and control. However, their modeling and optimization pose significant challenges to computing and storage resources, necessitating compact system representations via dimensionality reduction techniques, particularly the sparse and low-rank modeling paradigms. For multivariate stochastic systems, the sparse modeling emphasizes feature/variable selection through parameter sparsification, whereas low-rank modeling exploits intrinsic correlations among high-dimensional response variables to enable efficient data compression \cite{chen2012sparse}. The latter has demonstrated exceptional utility across diverse domains, including signal processing \cite{cai2010singular,candes2012exact,davenport2016overview,gu2014weighted,hu2012fast,Reweighted2019}, statistics  \cite{breiman1996heuristics,bunea2011optimal,chen2013reduced,izenman1975reduced,reinsel2022multivariate,yuan2007dimension,peng2024graph}, neural network compression  \cite{chen2021drone,chen2023joint,denil2013predicting,jaderberg2014speeding,papadimitriou2021data,tai2015convolutional,xu2019trained,yu2017compressing}, and systems and control \cite{cao2023identification,elisei2019recursive,fazel2004rank,fazel2001rank,fazel2013hankel,paleologu2018linear}. This work focuses on introducing online algorithms for the low-rank parameter matrix recovery of multivariate stochastic systems with non-stationary observations and establishing rigorous theoretical guarantees for their performance.

In signal processing, recovery of low-rank matrix aims to reconstruct an underlying matrix through the rank minimization or with a rank constraint by using the noisy or incomplete observations of the matrix and has wide applications in matrix completion, image processing, etc., \cite{davenport2016overview,gu2014weighted,Reweighted2019,cai2010singular,candes2012exact,negahban2011estimation}. The core challenge lies in the fact that minimizing the matrix rank (a non-convex, combinatorial objective) is NP-hard for general matrices. To address this, a foundational paradigm is to employ convex relaxations of the rank function. The most widely used approach is nuclear norm regularization (equivalent to the sum of singular values) \cite{cai2010singular}. Subsequent advancements have expanded this framework to include the adaptive weighted nuclear norm regularization and the non-convex alternatives such as the Schatten-$p$ norm \cite{lu2014generalized}, the minimax concave penalty (MCP) \cite{wen2018survey}, the truncated nuclear norm regularization (TNNR) \cite{hu2012fast}, etc. Theoretical analysis of these methods typically relies on the measurement operators of the underlying matrix to satisfy certain conditions such as the restricted isometry property (RIP) \cite{candes2011tight} or the restricted strong convexity (RSC) condition \cite{negahban2011estimation,negahban2009unified}, etc. These conditions are critical for establishing recovery guarantees. Complementary to regularization-based approaches are matrix factorization techniques, which decompose high-dimensional matrices into the product of low-dimensional factor matrices. Classic methods like singular value decomposition (SVD) provide a foundational framework, while generalized variants extend its applicability to different settings \cite{buchanan2005damped,eriksson2010efficient,ke2005robust}.

In multivariate statistical modeling, reduced rank regression (RRR) aims to estimate the coefficient matrix in the multivariate regression model with a key constraint that the coefficient matrix has a prescribed rank, which imposes a low-dimensional structure on the coefficient matrix to capture dominant variation patterns, reduce dimensionality, and improve noise resilience \cite{reinsel2022multivariate}. Under the normality assumption of the observations, the asymptotic distribution of the RRR estimator is established in \cite{izenman1975reduced}, showing that the estimator converges to the true low-rank matrix at a rate dependent on the signal strength and sample size. However, traditional RRR suffers from a limitation: the rank must be pre-specified {\em a priori}. Common strategies like likelihood ratio testing or cross-validation often exhibit instability, particularly in small-to-moderate sample sizes or when the true rank is near the upper bound of matrix dimension \cite{breiman1996heuristics}. Recent advancements have addressed the rank specification challenge through data-driven regularization frameworks that jointly estimate the rank and coefficient matrix, for example, the penalty of matrix rank \cite{bunea2011optimal}, the nuclear norm regularization \cite{yuan2007dimension}, the adaptive weighted nuclear norm regularization \cite{chen2013reduced}, etc. These methods guarantee consistent rank estimation with probability tending to one as the number of observations tending to infinity under the orthogonal design of the observations \cite{yuan2007dimension}, the signal-to-noise ratio condition of the model, the variance condition of noises and the Gaussian/sub-Gaussian noises \cite{bunea2011optimal,chen2013reduced}.

In neural network optimization, low-rank approximation (LRA) has emerged as a powerful compression technique for weight matrices, aiming to project high-dimensional parameters onto low-rank subspaces. This approach reduces computational complexity and storage requirements while preserving model performance \cite{denil2013predicting}. LRA can be applied in two paradigms: in-training compression (integrating rank constraints into the training objective) \cite{alvarez2017compression,xu2019trained} and post-training compression (applying low-rank factorization to pre-trained models) \cite{chen2021drone,chen2023joint,hsu2022language,li2023losparse,tai2015convolutional,wang2025svdllm,yuan2023asvd}. For example, for weight matrix compression of the trained large language model (LLM), notable methods include the activation-aware matrix singular value decomposition (SVD) \cite{yuan2023asvd}, the Fisher-weighted SVD \cite{hsu2022language}, the truncation-aware SVD \cite{wang2025svdllm}, the data-aware low-rank compression \cite{chen2021drone}, and the hybrid low-rank sparse approximation \cite{li2023losparse}. These methods have demonstrated remarkable practical success in applications. However, theoretical investigations are being pursued with increasing depth and methodological rigor.

Low-rank matrix recovery has emerged as a critical enabling technique in the field of systems and control, demonstrating its theoretical significance and practical applicability. For example, methodologies for matrix rank minimization under linear matrix inequality (LMI) constraints have evolved into useful tools for tackling challenges in control systems, including minimum-order controller synthesis, system realization theory, and data-driven system identification \cite{fazel2004rank,fazel2001rank,fazel2013hankel}, etc. Another line of research focuses on leveraging the low-rank properties of impulse response matrices through Kronecker product decompositions \cite{elisei2019recursive, dogariu2020efficient,paleologu2018linear}. Notably, in the context of linear stochastic systems, the work in \cite{li2022high} introduces a nuclear norm regularization criterion for low-rank matrix identification. Under the assumption that both observations and noise terms follow sub-Gaussian distributions, the study rigorously derives probability bounds on the estimation error, providing a theoretical foundation for robust state-space modeling in noisy environments.

In the context of system modeling, the existing literature mainly focuses on recovery of the low-rank matrix in static models \cite{davenport2016overview,gu2014weighted,Reweighted2019,cai2010singular,candes2012exact,negahban2011estimation} or dynamic models with sequential observations \cite{bunea2011optimal,chen2013reduced,yuan2007dimension}. Notably, in diverse fields such as systems and control, communication, and beyond, observation sequences often exhibit statistical non-stationarity, i.e., that is, the underlying random processes lack time-invariant probability distributions or statistics. Consequently, technical assumptions widely adopted in prior works such as the RIP/RSC conditions on the observation operator \cite{candes2011tight,negahban2011estimation,negahban2009unified}, the orthogonal observation designs \cite{yuan2007dimension}, the signal-to-noise ratio constraints, the noise variance conditions, or the Gaussian noise assumptions  \cite{bunea2011optimal,chen2013reduced}, may no longer hold. To the authors' knowledge, research on recovering the low-rank parameter matrices in multivariate stochastic systems with non-stationary observations remains scarce. This gap highlights a theoretical problem: traditional low-rank recovery frameworks, which rely heavily on stationary statistical observations or some other assumptions on observation operators, are ill-suited for non-stationary scenarios in real-world applications.

In this paper, we present a two-stage online algorithms for recovery of low-rank parameter matrix in non-stationary stochastic systems. The first stage employs a recursive least squares (RLS) estimator combined with its singular value decomposition (SVD) to estimate the unknown parameter matrix within the system. Specifically, RLS is leveraged for its adaptability to time-varying dynamics, while SVD is used to exploit the inherent low-rank structure of the parameter matrix. The second stage introduces a weighted nuclear norm regularization criterion function, where adaptive weights derived from the first-stage are incorporated to enhance the low-rank constraint. Notably, the regularization criterion admits an explicit and online computable solution, enabling efficient online updates when new data arrive without reprocessing historical data. Under the non-stationary and non-persistent excitation conditions on the systems, we establish the following theoretical results:
\begin{itemize}
\item[(i)] the true rank of the unknown parameter matrix can be identified with a finite number of observations,

\item[(ii)] the values of the matrix components can be consistently estimated as the number of observations increases,

\item[(iii)] the asymptotical normality of the algorithms is established as well,
\end{itemize}
for which the proposed algorithms are termed oracle properties in the literature.

The rest of the paper are organized as follows. The problem formulation and the online algorithms for recovering the low-rank parameter matrix are introduced in Section \ref{sec:proalg}. The theoretical results are established in Section \ref{sec:th}. In Section \ref{sec:simu} numerical simulations are given to validate performance of the algorithms in estimation accuracy. In Section \ref{sec:conclsn}, some concluding remarks are addressed.

\textbf{Notations.} Let $(\Omega, \mathcal{F}, \mathbb{P})$ be the probability space and $\omega$ be an element in $\Omega$. The mathematical expectation in $(\Omega, \mathcal{F}, \mathbb{P})$ is denoted by $\mathbb{E}(\cdot)$. The Frobenius norm of vectors and matrices is denoted by $\| \cdot \|$. For positive sequences $\{a_k\}_{k \geq 1}$ and $\{b_k\}_{k \geq 1}$, $a_k = O(b_k)$ means $a_k \leq cb_k,~k \geq 1$ for some $c > 0$ and $a_k = o(b_k)$ means $a_k/b_k \rightarrow 0$ as $k \rightarrow \infty$. For random sequences $\{x_k\}_{k \geq 1}$ and $\{y_k\}_{k \geq 1}$, $x_k = O_p(y_k)$ means $\{x_k\}_{k \geq 1}$ is bounded by $\{y_k\}_{k \geq 1}$ in probability. For a random sequence $\{x_k\}_{k \geq 1}$ and a random variable $x$, by $x_k\mathop{\longrightarrow}\limits^{\mathrm{a.s.}}x$, $x_k\mathop{\longrightarrow}\limits^{\textcolor{black}{\mathbb{P}}}x$, and $x_k\mathop{\longrightarrow}\limits^{\mathcal{D}}x$, we mean $\{x_k\}_{k\geq1}$ converge to $x$ almost surely, in probability, and in distribution, respectively. For a matrix $A$, denote its $\left(s,t\right)$-th entry by $A(s,t)$ and its $s$-th column by $A(s)$, while for a vector $v\in \mathbb{R}^m$, denote its $l$-th entry by $v(l)$. The maximal and minimal eigenvalues of a symmetric matrix $A$ are denoted by $\lambda_{\max}\{A\}$ and $\lambda_{\min}\{A\}$, respectively. Denote the $i$-th singular value of a matrix $A$ by $\sigma_i(A)$. The determinant and trace of a square matrix are denoted by ${\rm det}(\cdot)$ and $\mathrm{tr}\{\cdot\}$, respectively. The Kronecker operator of matrix product is denoted by $\otimes$. For a matrix $X\in \mathbb{R}^{d\times n}$ and a vector $w=[w_1,\cdots,w_n]^{\top},~w_i\geq0,~i=1,\cdots,n$, denote by $\|X\|_*$ the nuclear norm of $X$, i.e., $\|X\|_*=\sum_{i=1}^n\sigma_i(X)$ with $\sigma_i(X),~i=1,\cdots,n$ the singular values of $X$ and by $\|X\|_{w,*}=\sum_{i=1}^nw_i\sigma_i(X)$ the weighted sum of its singular values.

\section{Problem Formulation and Online Algorithms for Low-Rank Parameter Matrix Recovery}\label{sec:proalg}

Focusing on system identification, we will consider the recovery of the unknown low-rank parameter matrix $\Theta \in \mathbb{R}^{d \times n}$ of the following multivariate stochastic system,
\begin{align}\label{eq:linear}
    y_{k+1}=\Theta^{\top} \varphi_k + \varepsilon_{k+1},~k \geq 1,
\end{align}
where $\varphi_k \in \mathbb{R}^d$, $y_{k+1}\in \mathbb{R}^n$, and $\varepsilon_{k+1}\in \mathbb{R}^n$ are the regression vector, the output, and the noise, respectively. We assume that the process $\{\varphi_k,y_{k+1}\}_{k\geq1}$ is non-stationary. In this paper, by the non-stationarity, \textcolor{black}{it means that for any $k\geq1$,
\begin{align}
\varphi_k\in \mathcal{F}_k \triangleq \sigma \{\varphi_{k-1},\cdots,\varphi_1,{\varepsilon}_k,\cdots,{\varepsilon}_1,{\varepsilon}'_k,\cdots,{\varepsilon}'_1\},\label{eq2}
\end{align}}
where $\{{\varepsilon}'_k\}_{k\geq1}$ is a sequence of exogenous inputs or dithers to the system (\ref{eq:linear}). Notably, the system (\ref{eq:linear}) under the condition (\ref{eq2}) include the feedback control systems and the stationary time-series models as special cases.

Assume that $r=\mathrm{rank}\textcolor{black}{(\Theta)}\leq\min(d,n)$. Without loss of generality, we assume that $d\geq n$. Recovery of the low-rank parameter matrix $\Theta$ is to correctly identify the rank $r$ and to estimate the values of entries of $\Theta$, under the non-stationary condition (\ref{eq2}).
Additionally, the algorithms are required to have online computable solutions, which should enable efficient online updates when new data arrive  without the need to reprocess historical data.

\begin{remark}
The low-rank structural information of $\Theta$ facilitates system (\ref{eq:linear}) in various applications, such as controller synthesis, system realization, and system identification. For instance, leveraging the rank property, $\Theta$ can be decomposed into the product of two full-rank matrices $\Gamma\in \mathbb{R}^{d \times r}$ and $\Sigma\in \mathbb{R}^{n \times r}$, expressed as $\Theta=\Gamma\Sigma^{\top}$. This decomposition reduces storage requirements and enables lightweight system realization when $d r+n r<d n$ is satisfied. 
\end{remark}

Denote the singular value decomposition (SVD) of $\Theta$ by $\Theta=U \Sigma V^{\top}=U^{(1)} \Sigma^{(1)} {V^{(1)}}^{\top}$, where
$\Sigma=
\left[
\begin{smallmatrix}
    \Sigma^{(1)} & &0\\
    0& &0
\end{smallmatrix} \right]
$, $\Sigma^{(1)}={\rm diag}\{\sigma_1(\Theta),\cdots,\sigma_r(\Theta)\}$, the singular values of $\Theta$ are arranged in the descending order, i.e., $\sigma_1(\Theta)\geq \sigma_2(\Theta)\geq \cdots \geq \sigma_r(\Theta)$, $U \in \mathbb{R}^{d \times d}$ and $V \in \mathbb{R}^{n \times n}$ are both orthogonal matrices, and $U^{(1)}$ and $V^{(1)}$ represent the first $r$ columns of $U$ and $V$, respectively. The low-rank parameter matrix recovery algorithm in this work consists of two steps: the LS algorithm generating initial estimates for $\Theta$ and a weighted nuclear norm regularization criterion function estimating the matrix rank, where adaptive weights derived from the first-stage enhance low-rank constraints.

With the observations $\{\varphi_k,y_{k+1}\}_{N\geq1}$, denote the LS estimates by $\{\Theta_N\}_{N\geq1}$ and its SVD by $\Theta_{N+1}={U}_{N+1} {\Sigma}_{N+1} {V}_{N+1}^{\top}$, where
${\Sigma}_{N+1} \triangleq
\left[
\begin{smallmatrix}
    {\rm diag}\left\{\sigma_{1}\left(\Theta_{N+1}\right),\cdots,\sigma_{n}\left(\Theta_{N+1}\right)\right\}\\
    0_{(d-n)\times n}
\end{smallmatrix}
\right]$ and $\sigma_{1}\left(\Theta_{N+1}\right)\geq \sigma_{2}\left(\Theta_{N+1}\right) \geq \cdots \geq \sigma_{n}\left(\Theta_{N+1}\right)$.

For $i=1,\cdots n$, set
\begin{align}
&\widehat{\sigma}_{i}\left(\Theta_{N+1}\right) \triangleq \sigma_{i}\left(\Theta_{N+1}\right)+ \sqrt{\frac{\log\lambda_{\max}(N)}{\lambda_{\min}(N)}},\label{eq:shat}\\
&w_{N+1}(i)\triangleq \left(\widehat{\sigma}_{i}\left(\Theta_{N+1}\right)\right)^{-1},\label{eq4}
\end{align}
where ${{\lambda}}_{\max}(N)\triangleq \lambda_{\max}\{\sum_{k=1}^N \varphi_k \varphi_k^\top+\textcolor{black}{P_1}^{-1}\}$ and ${\lambda}_{\min}(N)\triangleq \lambda_{\min}\{\sum_{k=1}^N \varphi_k \varphi_k^\top+\textcolor{black}{P_1}^{-1}\}$. With the LS estimate $\Theta_{N+1}$ and the adaptive weights  $w_{N+1}(i),~i=1,\cdots,n$, estimates for the low-rank matrix are given by minimizing the following weighted nuclear norm regularization criterion,
\begin{align}
&J_{N+1}(X)\triangleq \frac{1}{2}\| \Theta_{N+1}-X \|^2 + \lambda_N \sum_{i=1}^n w_{N+1}(i)\sigma_i(X),\label{pro:J}\\
&X_{N+1}\triangleq\mathop{\arg \min}\limits_{X\in \mathbb{R}^{d\times n}} J_{N+1}(X),\label{eq6}
\end{align}
where $\left\{\lambda_N\right\}_{N\geq 1}$ is a positive sequence specified later.

Although $J_{N+1}(X)$ is non-convex in general due to its weighting terms $w_{N+1}(i),~i=1,\cdots,n$ (see, e.g., \cite{chen2013reduced}), for $X_{N+1}$ generated by (\ref{eq6}), the following result takes place.

\begin{lemma}\label{lem:JN}
Considering the minimization of the objective function (\ref{pro:J}), we have
\begin{align}\label{eq:defXN}
    X_{N+1}
    &\triangleq U_{N+1}
    \left[
    \begin{smallmatrix}
        {\rm diag}\left\{\max\left\{\sigma_i(\Theta_{N+1})-\lambda_N w_{N+1}(i), 0\right\}, i=1, \cdots, n\right\}\\
        0_{(d-n)\times n}
    \end{smallmatrix}
    \right]
    V_{N+1}^\top \nonumber\\
    &\in \mathop{\arg \min} \limits_{X \in \mathbb{R}^{d \times n}} J_{N+1}(X),
\end{align}
where $U_{N+1}$ and $V_{N+1}$ are the left and right singular matrices of the LS estimate $\Theta_{N+1}$, respectively.
\end{lemma}

\begin{IEEEproof}
The proof directly follows from Lemma \ref{lem:closedsol} given in the appendix. Please see Appendix B for details.
\end{IEEEproof}

Based on (\ref{eq:defXN}), the estimates for the matrix rank $r$ are given as follows:
\begin{align}\label{eq:rhat}
    \widehat{r}_N\triangleq \max \{i=1,\cdots,n |&1\leq j \leq i,\nonumber\\
    &\sigma_j\left(\Theta_{N+1}\right)\geq\lambda_N w_{N+1}(j)\}.
\end{align}

Then $X_{N+1}$ and $\widehat{r}_N$ generated by the above algorithms serve as estimates for $\Theta$ and its rank value $r$, respectively. 
Noting that in the above algorithms the LS algorithm can be recursively estimated and the weighted nuclear norm regularization criterion has a closed-form solution, the algorithms can be updated in an online manner efficiently when new data arrive without reprocessing historical data (see Algorithm \ref{Alg:1}).

\begin{algorithm}
\caption{: Online algorithms for low-rank matrix recovery}
\label{Alg:1}
\begin{algorithmic}
{\footnotesize
\STATE {\bf Initials:} $\mu>0,~P_1=\mu I_d,~\Theta_1=0,~\{\lambda_N\}_{N\geq 1}$ with $\lambda_N>0$\\
\textbf{for} $N=1,2,\cdots$ \textbf{ do }\\
\textbf{~~~Step 1: Recursive LS estimation and SVD for $\Theta$} \\
\STATE {
$~~~a_N = \left(1 + \varphi_N^\top P_N \varphi_N\right)^{-1}$\\
$~~~\Theta_{N+1}=\Theta_N+a_N P_N\varphi_N\left(y_{N+1}^\top-\varphi_N^\top \Theta_N\right)$\\
$~~~P_{N+1}=P_N-a_N P_N \varphi_N \varphi_N^\top P_N$}\\
\STATE {
$~~~\left[{U}_{N+1}, {\Sigma}_{N+1}, {V}_{N+1}\right]={\rm SVD}\left(\Theta_{N+1}\right)$
}
\STATE {~~~\textbf{for} $i=1,2,\cdots,n$, \textbf{ set }\\
$~~~\widehat{\sigma}_{i}\left(\Theta_{N+1}\right) = \sigma_{i}\left(\Theta_{N+1}\right)+ \sqrt{\frac{\log\lambda_{\max}(N)}{\lambda_{\min}(N)}}$\\
$~~~w_{N+1}(i)= \left(\widehat{\sigma}_{i}\left(\Theta_{N+1}\right)\right)^{-1}$
}\\
\textbf{~~~Step 2: Recovery of low-rank matrix $\Theta$}
\STATE{
$~~~
X_{N+1}$\\
$~~~=U_{N+1} \left[
    \begin{smallmatrix}
        {\rm diag}\left\{\max\left\{\sigma_i(\Theta_{N+1})-\lambda_N w_{N+1}(i), 0\right\}, i=1, \cdots, n\right\}\\
        0_{(d-n)\times n}
    \end{smallmatrix}
    \right] V_{N+1}^\top
$\\
$~~~\widehat{r}_N = \max \left\{i=1,\cdots,n~\left|~1\leq j \leq i,~\sigma_j\left(\Theta_{N+1}\right)\geq\lambda_N w_{N+1}(j)\right.\right\}$
}\\
\textbf{end for;}
}
\end{algorithmic}
\end{algorithm}

\begin{remark}
In Algorithm \ref{Alg:1}, the LS algorithm provides initial estimates for $\Theta$. In fact, other algorithms such as the stochastic gradient (SG) algorithm and the instrumental variable (IV) algorithm, can also be applied in the first stage.
In Algorithm \ref{Alg:1}, SVD is also involved which can be efficiently solved by using numerical methods, such as the eigenvalue decomposition \cite{stewart2001matrix}, the QR decomposition \cite{golub2013matrix}, the iterative method \cite{saad2003iterative}, and the randomized method \cite{halko2011finding}, etc.
\end{remark}

\section{Theoretical Properties of Algorithms}\label{sec:th}

We introduce the following assumptions.

\begin{assumption}\label{assum:1}
The system noise $\{\varepsilon_k,\mathcal{F}_k\}_{k \geq 1}$ is a martingale difference sequence ({\rm m.d.s.}), i.e., $\mathbb{E}\left[\varepsilon_{k+1}| \mathcal{F}_k\right]=0\in\mathbb{R}^n,k \geq 1,$ and there exists some $\beta > 2$ such that ${\rm \sup}_k \mathbb{E}[\|\varepsilon_{k+1}\|^{\beta}| \mathcal{F}_k]< \infty,~~\mathrm{a.s.}$
\end{assumption}
\begin{assumption}\label{assum:2}
For each $k \geq 1$, $\varphi_k$ is $\mathcal{F}_k$-measurable.
\end{assumption}
\begin{assumption}\label{assum:3}
For the maximal and minimal eigenvalues of $\sum_{k=1}^{N} \varphi_k \varphi_k^{\top}+\frac{1}{\mu} I_d$ with $\mu\in(0,1)$, it holds that
\begin{align}
\frac{\log\lambda_{\max}(N)}{\lambda_{\min}(N)} \mathop{\longrightarrow}\limits_{N \rightarrow \infty} 0, ~~\mathrm{a.s.}\label{eq9}
\end{align}
\end{assumption}
\begin{assumption}\label{assum:lambdan01}
The regularization coefficient $\left\{\lambda_{N}\right\}_{N\geq 1}$ with $\lambda_N>0$ satisfies
\begin{align}\label{eq:lambdan01}
\lambda_N=O\left(\sqrt{\frac{\log\lambda_{\max}(N)}{\lambda_{\min}(N)}}\right),~\frac{\log\lambda_{\max}(N)}{\lambda_{\min}(N)}=o(\lambda_N),~~{\rm a.s.}
\end{align}
\end{assumption}

\begin{remark}
Assumption \ref{assum:3} is a relaxed excitation condition for identification of system (\ref{eq:linear}). It is direct to check that (\ref{eq9}) includes the classical persistent excitation condition, which indicates $\lambda_{\max}(N)\sim N$ and $\lambda_{\min}(N)\sim N$, as a special case.
\end{remark}

\subsection{Convergence of Estimates for Matrix Rank}\label{subsec:rank}

For the rank estimates $\widehat{r}_N$ generated by (\ref{eq:rhat}), we have the following result.

\begin{theorem}\label{th:rank}
Assume that Assumptions \ref{assum:1}, \ref{assum:2}, \ref{assum:3}, and \ref{assum:lambdan01} hold. Then there exists an $\omega$-set $\Omega_0$ with $\mathbb{P}\{\Omega_0\}=1$ and for any $\omega \in \Omega_0$, there exists an integer $N_0(\omega)>0$ such that on the trajectory $\{\widehat{r}_{N}(\omega)\}_{N\geq1}$ generated by $\omega$,
\begin{align}
\widehat{r}_{N}=\mathrm{rank}(\Theta)=r,~N \geq N_0(\omega),\label{eq11}
\end{align}
and
\begin{align}
\mathrm{rank}(X_{N+1})=\mathrm{rank}(\Theta)=r,~N \geq N_0(\omega).\label{eq12}
\end{align}
\end{theorem}

\begin{IEEEproof}
The basic idea of the proof is to compare the magnitudes of the singular values $\sigma_i(\Theta_{N+1})$ and the threshold values $\lambda_N w_{N+1}(i)$ for $i\in \{1,\cdots,n\}$ through the choice of $\lambda_N$ and the use of the asymptotic error bounds on $\|\Theta_{N+1}-\Theta\|$. Please see Appendix C for details.
\end{IEEEproof}

\subsection{Convergence of Estimates for Matrix Entries}\label{subsec:para}

The following theorem establishes the asymptotic error bound and the almost sure convergence of $\{X_{N}\}_{N\geq1}$ generated by Algorithm \ref{Alg:1}.

\begin{theorem}\label{th:errorbound}
Assume that Assumptions \ref{assum:1}, \ref{assum:2}, \ref{assum:3}, and \ref{assum:lambdan01} hold. Then
\begin{align}
\|X_{N+1}-\Theta\|=O\left(\sqrt{\frac{\log \lambda_{\max}(N)}{\lambda_{\min}(N)}}\right),~~ \mathrm{a.s.}
\end{align}
and
$X_{N+1}(s,t) \mathop{\rightarrow} \Theta(s,t)$ as $N \rightarrow \infty$ for $s=1,\cdots,d,~t=1,\cdots,n\textcolor{black}{,~\mathrm{a.s.}}$
\end{theorem}

\begin{IEEEproof}
The proof is based on the consistency in rank estimation established in Theorem \ref{th:rank} and the error bound of the LS algorithm in Lemma \ref{lem:lserr}. Please see Appendix D for details.
\end{IEEEproof}

\begin{remark}
Theorems \ref{th:rank} and \ref{th:errorbound} show that $X_{N+1}$ generated by (\ref{eq:defXN}) provides consistent estimates for the rank and the values of entries of the low-rank matrix $\Theta$.
\end{remark}

\subsection{Asymptotic Normality of Estimates}\label{subsec:asy}

To analyze the asymptotic distribution of $\{X_{N+1}\}_{N\geq1}$, we further introduce the following assumptions.

\begin{assumption}\label{assum:4}
$\{\varepsilon_k,\mathcal{F}_k\}_{k \geq 1}$ satisfies Assumption \ref{assum:1} and $\mathbb{E}\left[\varepsilon_{k+1} \varepsilon_{k+1}^\top| \mathcal{F}_k\right]=\sigma^2 I_n,~~\mathrm{a.s.}$ for some $0<\sigma^2 < \infty$.
\end{assumption}

\begin{assumption}\label{assum:5}
There exists a sequence of deterministic positive definite matrices $\left\{B_N\right\}_{N\geq 1}$ such that
\begin{align}
B_N^{-1} \left(\sum_{k=1}^N \varphi_k \varphi_k^\top\right)^{1/2} \mathop{\longrightarrow} \limits^{\mathbb{P}}_{N \rightarrow \infty} I_d,~\max _{1 \leq k \leq N}\left\|B_N^{-1} \varphi_k\right\| \mathop{\longrightarrow} \limits^{\mathbb{P}}_{N \rightarrow \infty} 0.
\end{align}
\end{assumption}

\begin{assumption}\label{assum:6}
i) \textcolor{black}{If $\Theta$ is of low rank, i.e., ${\rm rank}(\Theta)=r<n$,} there exists a sequence of matrices $\left\{C_{N}\right\}_{N\geq 1}$ and a constant matrix $M$ for which
\begin{align}
C_{N} U^{(1)} {U^{(1)}}^\top \left(\sum_{k=1}^N \varphi_k \varphi_k^\top\right)^{-1/2} \mathop{\longrightarrow} \limits^{\mathbb{P}}_{N \rightarrow \infty} M,~\lambda_N \left\|C_N\right\| \mathop{\longrightarrow} \limits^{\mathbb{P}}_{N \rightarrow \infty} 0,\label{eq:cnlambdan}
\end{align}
and
\begin{align}\label{eq:cnrate}
\frac{\sqrt{\log\lambda_{\max}(N)}}{\lambda_{\min}(N)}\left\|C_{N}\right\|\mathop{\longrightarrow} \limits^{\mathbb{P}}_{N \rightarrow \infty} 0,
\end{align}
where $U^{(1)}$ is a matrix with the first $r$ columns of the left singular matrix of $\Theta$ and $\left\{\lambda_{N}\right\}_{N\geq 1}$ satisfies Assumption \ref{assum:lambdan01}.

ii) \textcolor{black}{If $\Theta$ is of full rank, i.e., ${\rm rank}(\Theta)=n$, there exists a sequence of matrices $\left\{C_{N}\right\}_{N\geq 1}$ for which
\begin{align}\label{eq:cnlambdanfull}
C_{N} \left(\sum_{k=1}^N \varphi_k \varphi_k^\top\right)^{-1/2} \mathop{\longrightarrow} \limits^{\mathbb{P}}_{N \rightarrow \infty} I_d,~~\lambda_N \left\|C_N\right\| \mathop{\longrightarrow} \limits^{\mathbb{P}}_{N \rightarrow \infty} 0,
\end{align}
where $\left\{\lambda_{N}\right\}_{N\geq 1}$ satisfies Assumption \ref{assum:lambdan01}.}
\end{assumption}
\begin{assumption}\label{assum:theta}
The nonzero singular values of the true parameter matrix $\Theta$ are pairwise distinct.
\end{assumption}

\begin{remark}
Conditions similar to Assumption \ref{assum:5} are applied in \cite{lai1981consistency,zhang2023adaptive}. Here we give an example for this assumption.
If $\{\varphi_k \}_{k\geq1}$ is bounded and stationary with $\mathbb{E}\left[\varphi_1 \varphi_1^\top\right]>0$, by setting $B_N=\sqrt{N}\left\{\mathbb{E}\left[\varphi_1 \varphi_1^\top\right]\right\}^{1/2}$, we have $\max _{1 \leq k \leq N}\big\|B_N^{-1} \varphi_k\big\|
=\max _{1 \leq k \leq N}\big\|{N}^{-1/2}\left\{\mathbb{E}\left[\varphi_1 \varphi_1^\top\right]\right\}^{-1/2} \varphi_k\big\|
=O_p\big({N}^{-1/2}\big)
\mathop{\longrightarrow} \limits^{\mathbb{P}}_{N \rightarrow \infty} 0$ and Assumption \ref{assum:5} is satisfied.
\end{remark}

\begin{remark}\label{re:cn}
For Assumption \ref{assum:6}, we consider the following two examples.
\begin{sloppypar}
i) Assume that $\{\varphi_k \}_{k\geq1}$ is stationary and ergodic with $\mathbb{E}\left[\varphi_1 \varphi_1^\top\right]>0$. We set $C_N=\sqrt{N}\left\{\mathbb{E}\left[\varphi_1 \varphi_1^\top\right]\right\}^{1/2}$, $M=\left\{\mathbb{E}\left[\varphi_1 \varphi_1^\top\right]\right\}^{1/2} U^{(1)} {U^{(1)}}^\top \left\{\mathbb{E}\left[\varphi_1 \varphi_1^\top\right]\right\}^{-1/2}$ with ${\rm rank} (M)=r$, and $\lambda_N=\left(\frac{\log\lambda_{\max}(N)}{\lambda_{\min}(N)}\right)^{\left(1/2+\varepsilon\right)}$ with $0<\varepsilon<1/2$. It is direct to check that for the given $\{\lambda_N\}_{N\geq1}$, Assumption \ref{assum:3} is satisfied and for the given $\{\varphi_k \}_{k\geq1}$, the maximal and minimal eigenvalues of $\sum_{k=1}^N\varphi_k \varphi_k^\top$ satisfy that $\lambda_{\max}(N)\leq C_1N$ and $\lambda_{\min}(N)\geq C_2N$ for some constants $C_1,C_2>0$ which values may depend on the sample paths. Thus,
\begin{align*}
\lambda_N\|C_N\| &=O_p\left(\left(\frac{\log N}{N}\right)^{1/2+\varepsilon} \sqrt{N}\right)
=o_p(1),
\end{align*}
\textcolor{black}{
\begin{align*}
\frac{\sqrt{\log\lambda_{\max}(N)}}{\lambda_{\min}(N)}\left\|C_{N}\right\|
&=O_p\left(\frac{\sqrt{\log N}}{N} \sqrt{N}\right)
=o_p(1),
\end{align*}}
and Assumption \ref{assum:6} is satisfied.
\end{sloppypar}
ii) Assume that $\left\{\varphi_k\right\}_{k\geq 1}$ is a sequence of $d$-dimensional mutually independent and zero-mean Gaussian random vectors with covariance matrices $\mathbb{E}[\varphi_k \varphi_k^\top] = k^{\delta}\Sigma$ with $\Sigma$ being a positive definite matrix and $1>\delta > 0$. Clearly, $\left\{\varphi_k\right\}_{k\geq 1}$ is non-stationary since the distribution functions are time-varying. First by the Borel-Cantelli's lemma and then the Kronecker's lemma, we can prove that
\begin{align}\label{eq:nonsta}
\frac{1}{N^{\delta+1}} \sum_{k=1}^N \varphi_k \varphi_k^\top \mathop{\longrightarrow} \limits^{\mathrm{a.s.}}_{N \rightarrow \infty} \frac{\Sigma}{\delta+1}
\end{align}
and thus
\begin{align}\label{eq:nonsta'}
C_3 N^{\delta+1} \leq \lambda_{\min}(N) ,~~\lambda_{\max}(N) \leq C_4 N^{\delta+1},~~\mathrm{a.s.}
\end{align}
for some $C_3,C_4>0$ which may depend on the sample paths. Set $C_N = \big(\frac{N^{\delta+1}}{\delta+1} \Sigma\big)^{1/2}$ and $\lambda_N=\left(\frac{\log\lambda_{\max}(N)}{\lambda_{\min}(N)}\right)^{\left(1/2+\varepsilon\right)}$ with $0<\varepsilon<1/2$. By (\ref{eq:nonsta}) and (\ref{eq:nonsta'}) we can prove that
$$
C_{N} U^{(1)} {U^{(1)}}^\top \left(\sum_{k=1}^N \varphi_k \varphi_k^\top\right)^{-1/2} \mathop{\longrightarrow} \limits^{\mathbb{P}}_{N \rightarrow \infty} \Sigma^{1/2} U^{(1)} {U^{(1)}}^\top \Sigma^{-1/2},
$$
\begin{align}
\lambda_N\|C_N\|
&=O_p\left(\left(\frac{\log N^{\delta+1}}{N^{\delta+1}}\right)^{1/2+\varepsilon} N^{(\delta+1)/2}\right)\nonumber\\
&=O_p\left(\frac{\left(\log N\right)^{1/2+\varepsilon}}{N^{(\delta+1)\varepsilon}}\right)=o_p(1)\nonumber,
\end{align}
\textcolor{black}{
and
\begin{align*}
\frac{\sqrt{\log\lambda_{\max}(N)}}{\lambda_{\min}(N)}\left\|C_{N}\right\|
&=O_p\left(\frac{\sqrt{\log N^{\delta+1}}}{N^{\delta+1}} N^{(\delta+1)/2}\right)
=o_p(1).
\end{align*}
}
Assumption \ref{assum:6} is satisfied for this example.
\end{remark}

We first have two technical lemmas.

\begin{lemma}\label{lem:u1u1t}
Assume that Assumptions \ref{assum:1}, \ref{assum:2}, and \ref{assum:theta} hold. Then
\begin{align}
\left\|U_{N+1}^{(1)} {U_{N+1}^{(1)}}^\top - U^{(1)} {U^{(1)}}^\top\right\|=O\left(\sqrt{\frac{\log \lambda_{\max}(N)}{\lambda_{\min}(N)}}\right),~~{\rm a.s.}
\end{align}
where $U_{N+1}^{(1)}$ and $U^{(1)}$ are the first $r$ columns of the left singular matrix of $X_{N+1}$ and $\Theta$, respectively.
\end{lemma}

\begin{IEEEproof}
Please see Appendix E for details.
\end{IEEEproof}

\begin{lemma}\label{lem:LSasy}
Assume that Assumptions \ref{assum:1}, \ref{assum:2}, \ref{assum:4}, and \ref{assum:5} hold. Then
\begin{align}
{\rm vec}\left(\left(\sum_{k=1}^N \varphi_k \varphi_k^\top\right)^{-1/2} \sum_{k=1}^N \varphi_k \varepsilon_{k+1}^\top\right) \mathop{\longrightarrow} \limits^\mathcal{D}_{N \rightarrow \infty} \mathcal{N}\left(0_{dn}, \sigma^2 I_{dn}\right).\label{eq19}
\end{align}
\end{lemma}

\begin{IEEEproof}
The result can be established by application of Lemma \ref{lem:clt} given in the appendix. Please see Appendix F for details.
\end{IEEEproof}

For the asymptotical distribution of Algorithm \ref{Alg:1}, we have the following result.

\begin{theorem}\label{th:asydisx}
\textcolor{black}{If ${\rm rank}(\Theta)=r<n$ and Assumptions \ref{assum:1}, \ref{assum:2}, \ref{assum:3}, \ref{assum:lambdan01}, \ref{assum:4}, \ref{assum:5}, \ref{assum:6}, and \ref{assum:theta} hold, then} $\{X_{N+1}\}_{N\geq1}$ generated by Algorithm \ref{Alg:1} asymptotically has a normal distribution in the sense that
\begin{align}\label{eq:asydiseq}
\left(I_n \otimes C_{N} U_{N+1}^{(1)} {U_{N+1}^{(1)}}^\top\right) {\rm vec}\left(X_{N+1}-\Theta\right)\nonumber\\
\mathop{\longrightarrow} \limits_{N \rightarrow \infty}^\mathcal{D} \mathcal{N}(0_{dn}, \sigma^2 I_n \otimes  M M^\top).
\end{align}
\textcolor{black}{If ${\rm rank}(\Theta)=n$ and Assumptions \ref{assum:1}, \ref{assum:2}, \ref{assum:4}, \ref{assum:5}, and \ref{assum:6} hold, then
\begin{align}\label{eq:asydiseq'}
\left(I_n \otimes C_{N}\right) {\rm vec}\left(X_{N+1}-\Theta\right)
\mathop{\longrightarrow} \limits_{N \rightarrow \infty}^\mathcal{D} \mathcal{N}(0_{dn}, \sigma^2 I_{dn}).
\end{align}}
\end{theorem}

\begin{IEEEproof}
Set $\Delta X_{N+1}\triangleq X_{N+1}-\Theta_{N+1}$. With the definition, it follows that
\begin{align}\label{eq:asyintro0}
&{U_{N+1}^{(1)}}^\top \left(X_{N+1}-\Theta\right) \nonumber\\
&={U_{N+1}^{(1)}}^\top P_{N+1} \sum_{k=1}^N \varphi_k \varepsilon_{k+1}^\top +{U_{N+1}^{(1)}}^\top P_{N+1}P_1^{-1}\left(\Theta_1-\Theta\right)\nonumber\\
&~~~-{U_{N+1}^{(1)}}^\top \Delta X_{N+1}.
\end{align}
The result can be obtained by analyzing the asymptotical distributions of the terms on the right side of (\ref{eq:asyintro0}). Please see Appendix G for details.
\end{IEEEproof}

\begin{remark}
\textcolor{black}{If ${\rm rank}(\Theta)=n$, by choosing $C_N=(\sum_{k=1}^N\varphi_k\varphi_k^{\top})^{1/2}$,} Theorem \ref{th:asydisx} indicates that
\begin{align}\label{eq:asydiseq1}
\Big(I_n \otimes \Big(\sum\limits_{k=1}^N\varphi_k\varphi_k^{\top}\Big)^{1/2} \Big) {\rm vec}\left(X_{N+1}\!\!-\!\Theta\right)
\mathop{\longrightarrow} \limits_{N \rightarrow \infty}^\mathcal{D} \mathcal{N}\left(0_{dn}, \sigma^2 I_{dn} \right)
\end{align}
provided the regularization coefficient $\lambda_N$ satisfying Assumption \ref{assum:6}. If $n=1$, i.e., $\Theta$ is a parameter vector, then the asymptotical distribution given in (\ref{eq:asydiseq1}) exactly coincides with the classical result of asymptotical distribution of the LS algorithm with non-stationary observations (Theorem 3 in \cite{lai1982least}). In this regard, Theorem \ref{th:asydisx} establishes the asymptotical distribution of the low-rank matrix recovery algorithms, extending the corresponding results of the LS algorithm.
\end{remark}

\section{Numerical Simulations}\label{sec:simu}

\newcommand\realset{\mathbb{R}}
\newcommand\complexset{\mathbb{C}}
\newcommand\complexnormal{\mathcal{CN}}
\newcommand\normal{\mathcal{N}}
\newcommand\rank{{\rm rank}}
\newcommand\channel{{\mathbf{H}}}
\newcommand\channelsub{{\mathbf{h}}}
\newcommand\mathif{\text{if }}
\newcommand\datas{\mathbf{ds}}
\newcommand{\steering}{{\mathbf{a}}}

We verify the low-rank matrix recovery algorithm by using data from a closed-loop control system and a wireless communication data set. The simulation is performed with MATLAB R2024a on a desktop with an Intel Core i7-10700 CPU and 16GB RAM.

{\em Example 1) } We first generate non-stationary data from a linear system with a self-tuning regulator (STR), which is a classical closed-loop adaptive control system (see, e.g., \cite{aastrom1973self}). We consider a multivariate stochastic system
$$
y_{k+1}+A_1 y_k +A_2y_{k-1}=B_1 u_k+B_2 u_{k-1}+w_{k+1},~k \geq 1
$$
where $A_1=-1.7\times I_{10},~A_2=0.7\times I_{10},~B_1= I_{10},~B_2=0.5\times I_{10}$ and $I_{10}$ is the identity matrix in $\mathbb{R}^{10 \times 10}$. The noise $\{w_k\}_{k\geq 1}$ is supposed to be i.i.d. with Gaussian distribution $\mathcal{N}(0,0.25\times I_{10})$. Set the reference signals $\{y_k^*\}_{k\geq 1},~y_k^*\in \mathbb{R}^{10}$, for any $i \in \{1,\cdots,10\}$,
$$
y_k^*(i)=
\left\{
\begin{aligned}
10,~&k \in [1000l+1,\cdots,1000l+500]\\
-10,~&k \in [1000l+501,\cdots,1000l+1000]
\end{aligned}
,~l \geq 0
\right..
$$
\begin{sloppypar}
Denote $\Theta^0\triangleq[-A_1,-A_2,B_1,B_2]^\top\in \mathbb{R}^{40 \times 10}$ and its LS estimates by $\Theta_k\triangleq \left[-A_{1,k},-A_{2,k},B_{1,k},B_{2,k}\right]^\top,~k \geq 1$. The self-tuning regulation control is given by (see, e.g., \cite{chen2012identification}),
$$u_k=u_k^0+\frac{\epsilon_k}{r_{k-1}^{\bar{\epsilon}/2}}, ~k \geq 1$$
where
$u_k^0=B_{1,k}^{-1}(y_{k+1}^*+B_{1,k} u_k -\Theta_k^\top \varphi_k)$,
$r_{k-1}=1+\sum_{i=1}^{k-1}\|\varphi_i\|^2$, and $\{\epsilon_k\}_{k\geq1}$ is an i.i.d. sequence with a uniform distribution over $[-0.1, 0.1]$ and is independent of $\{w_k\}_{k\geq1}$. In this example we choose $\bar{\epsilon} = 1/50$ in $\frac{\epsilon_k}{r_{k-1}^{\bar{\epsilon}/2}}$. Then we have the observations $\varphi_k=[y_k^{\top},y_{k-1}^{\top},u_k^{\top},u_{k-1}^{\top}]^\top \in  \mathbb{R}^{40},~k\geq1$.
\end{sloppypar}

Choose two matrices $\theta_1\in \mathbb{R}^{40\times 4}$ and $\theta_2\in \mathbb{R}^{4\times 40}$ with each component sampled from Gaussian distribution $\mathcal{N}\left(0,4\right)$. Set the true low-rank matrix $\Theta=\theta_1 \theta_2 \in \mathbb{R}^{40 \times 40}$ and consider the low-rank matrix recovery of the following system
$$
h_{k+1}=\Theta^{\top} \varphi_k + \varepsilon_{k+1},~k \geq 1,
$$
where $\{\varphi_{k}\}_{k \geq 1}$ are generated from STR and $\{\varepsilon_{k}\}_{k \geq 1}$ are independent and identically distributed with the normal distribution $\varepsilon_k \sim \mathcal{N}\left(0, 0.5 \times I_{40}\right)$. Set the regularization coefficients as $\lambda_N=N^{\alpha}$ with $\alpha=-0.1$ by comparing the test errors for $\alpha=-0.1,-0.2,-0.3,-0.4,-0.5$. In this example, we do 20 independent simulations, i.e., $T=20$, each with data length $N=4000$.

To evaluate the performance of the algorithm, we consider the following two indexes: the mean of relative parameter estimation error
$$\textbf{ParaEstErr}(X_N)\triangleq \frac{1}{T}\sum_{j=1}^T \|X_{j,N}-\Theta\|/\left\|\Theta\right\|$$
and the mean of rank estimation error
$$\textbf{RankEstError}(X_N) = \frac{1}{T}\sum_{j=1}^T |{\rm rank}(X_{j,N}) - {\rm rank}(\Theta)|$$ where $X_{j,N}$ is the estimate generated by the $j$-th simulation. Table \ref{Tab:2} compares the performance of the RLS algorithm and Algorithm \ref{Alg:1}.

\begin{table}
\caption{\textbf{ParaEstErr} and \textbf{RankEstError} in 20 simulations with data length $N=4000$.}
\label{Tab:2}
\begin{center}
\begin{tabular}{ccc}
\hline
\bfseries  & \textbf{ParaEstErr} & \textbf{RankEstError}\\
\hline
RLS & $9.3653\times {10}^{-4}$ & 36 \\
Algorithm \ref{Alg:1} & ${\bf 4.5516\times {10}^{-4}}$ & \textbf{0} \\
\hline
\end{tabular}
\end{center}
\end{table}

{\em Example 2)} Channel estimation is a crucial process in wireless communication, which is to understand the effects of the wireless channel, such as fading, scattering, reflection, and noises, so that the receiver can accurately reconstruct the transmitted signals \cite{chizhik2002keyholes}. Nowadays, large-scale multiple-input multiple-output (MIMO) systems are seen as a promising model to increase the speed of wireless communication. The transmission channel in MIMO systems typically exhibits low-rank characteristic \cite{chizhik2002keyholes}.

In wireless communication, channel estimation takes place before data transmission begins. First, the transmitter and receiver agree on a pilot signal $x_k$, which is known to them both. In this example, we choose the video signals\footnote{\url{http://cseweb.ucsd.edu/~viscomp/projects/LF/papers/ECCV20/nerf/website_renders/depth_ornament.mp4}} as the pilot signals, which are non-stationary. This video is 3MB in size, with resolution of $2016 \times 768$ and 30 frames per second.

\begin{itemize}
\item We firstly load the video into binary form, producing a binary data stream $\datas = 01001100101 \cdots$. Then, we map the binary data to complex symbol using quadrature amplitude modulation (QAM)
    $$
    x_{k}(i) = \frac{1}{\sqrt{d}}
    \begin{cases}
    -1 + j, & \mathif \datas_{[2i:2i+1]} = 00, \\
    -1 - j, & \mathif \datas_{[2i:2i+1]} = 01, \\
    1 + j, & \mathif \datas_{[2i:2i+1]} = 10, \\
    1 - j, & \mathif \datas_{[2i:2i+1]} = 11.
    \end{cases}
    $$

\item The channel matrix is given by
$$
\channel = \left[
    \begin{matrix}
        \channelsub_1, \channelsub_2, \cdots , \channelsub_n
    \end{matrix}
    \right]\in \mathbb{C}^{d\times n},
$$
with $\channelsub_i\in \mathbb{C}^d$ given by
$$
\channelsub_i =
\sqrt{\frac{d}{L}} \sum_{l=1}^L g_l e^{-j k_i r_l}  \steering(\theta_l),
$$
where $k_i = \frac{2 \pi f_i}{c}$ denotes the wavenumber, $c$ is the speed of light, $L$ is the number of paths in a multipath channel, $g_l, r_l,$ and $\theta_l$ denotes the complex path gain, the distance, and the angle of the $l$-th path, respectively. The steering vector $\steering(\theta_l)$ on angle $\theta_l$ is defined under planar-wave assumption \cite{cui2022channel}:
$$
\steering(\theta_l) = \frac{1}{\sqrt{d}}
\left[1, e^{j \pi \theta_l}, \cdots, e^{j (d-1)\pi \theta_l}\right]^\top,
$$
and the distances of paths are set as $r_1=\hdots=r_l=\hdots=r_L$. Therefore, we have
$
\channelsub_i =
\sqrt{\frac{d}{L}} e^{-j k_i r_L} \sum_{l=1}^L g_l  \steering(\theta_l)
$
which implies ${\rm rank}(\channel)=1$.
\end{itemize}

The channel can be modeled by
$$
y_k = \channel^{*} x_k + \varepsilon_k,~k\geq 1,
$$
where $\channel \in \complexset^{d \times n}$ is the channel matrix, $\channel^*$ is the conjugate transposition of $\channel$, $x_k \in \complexset^{d}$ denotes the transmitted signal, $y_k \in \complexset^{n}$ is the received signal, and $\varepsilon_k$ is the noise, typically assumed to be complex Gaussian distributed $\complexnormal \left(0, \frac1{\rm SNR} I_n \right)$, where ${\rm SNR}$ is the signal-to-noise ratio.

Set $d = 256$, $n = 16$, and ${\rm SNR} = 10 {\rm dB}$. With the observations $\{x_k,y_k\}_{k=1}^N$, the receiver estimates the channel matrix, resulting an estimate $\widehat{\channel}$. In this simulation, we do 20 independent simulations, i.e., $T = 20$. To evaluate the quality of channel estimates, in addition to the estimation error considered in Example 1), we also consider the normalised mean square error (NMSE) index
\begin{align}
    &{\bf NMSE} = \frac{1}{T}\sum_{j=1}^T \|\widehat{\channel}_{j,N} - \channel\|^2/\|\channel\|^2,\nonumber\\
    &{\bf NMSE_{dB}} = 10 \log_{10} {\bf NMSE}\nonumber
\end{align}
where $\widehat{\channel}_{j,N}$ is the estimate generated from the $j$-th simulation.
In this example, we set $N = 64, 128, 256, 512, 768$ to demonstrate the effectiveness of the proposed algorithm under different data length.

As a comparison, two large-scale MIMO channel estimation methods, i.e., the polar-domain simultaneous gridless weighted (P-SIGW) algorithm \cite{cui2022channel} and the oracle least squares (Oracle LS) \cite{cui2022channel} algorithm, are selected as baselines to compare the performance of Algorithm \ref{Alg:1}. Note that Algorithm \ref{Alg:1} is a two-stage algorithm. In the first stage, the role of the RLS algorithm is to provide an initial estimate for the unknown parameter matrix while other algorithms can also be applied in this stage.  In this example, we compare the performace of P-SIGW, Oracle LS, Algorithm \ref{Alg:1}, Algorithm \ref{Alg:1} with P-SIGW in the first stage, and Algorithm \ref{Alg:1} with Oracle LS in the first stage.

\begin{figure}[tbhp]
    \centering
    \includegraphics[width=8cm]{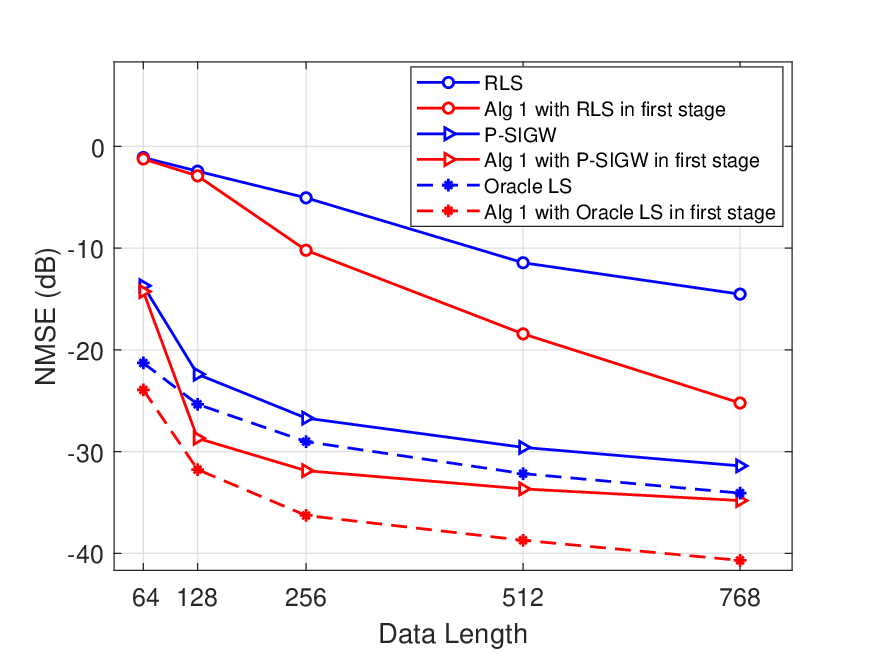}
    \includegraphics[width=8cm]{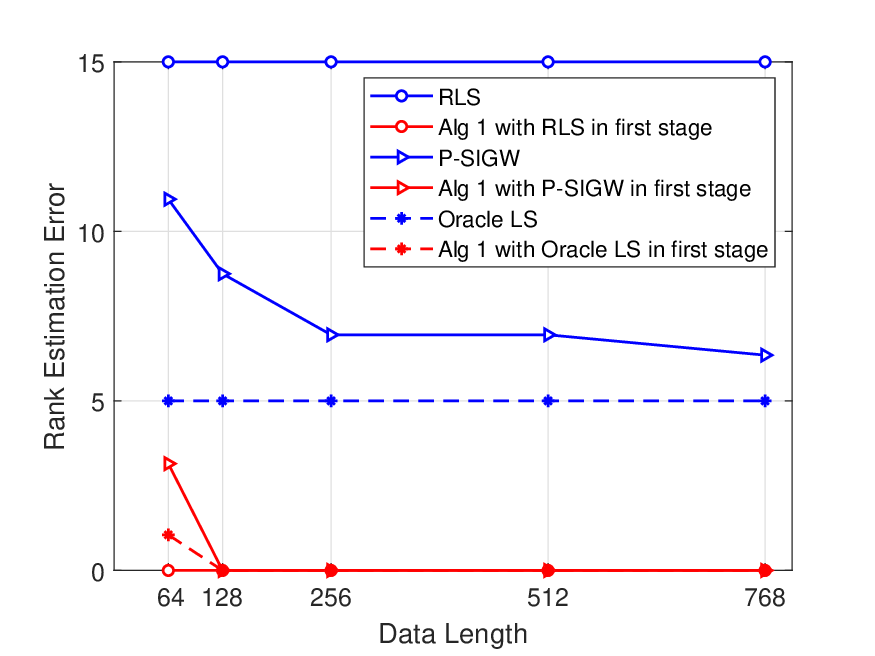}
    \caption{ ${\bf NMSE_{dB}}$ and \textbf{RankEstError} of channel estimation generated by RLS, Algorithm \ref{Alg:1}, P-SIGW, Algorithm \ref{Alg:1} with P-SIGW in first stage, Oracle LS, and Algorithm \ref{Alg:1} with Oracle LS in first stage, respectively. }
    \label{Fig:channelest}
\end{figure}

For Algorithm \ref{Alg:1}, we choose the regularization coefficient $\lambda_N=N^{\alpha}$ with $\alpha=0.5$ by comparing the performance of $\alpha=-0.9,-0.7,-0.5,-0.3,-0.1,0.1,0.3,0.5,0.7,0.9$. Fig \ref{Fig:channelest} shows the performance of the algorithms, from which we find that to recover the low-rank property of the channel matrix significantly improves the mean square error of the channel modeling.


\section{Conclusions}\label{sec:conclsn}

This paper presents an online algorithm for low-rank matrix identification of multivariate linear stochastic systems, which employs nuclear norm regularization with adaptive weights. Under non-stationary signals, we theoretically establish the correctness of rank inference, strong consistency in estimation of values of matrix components, and asymptotic normality of the proposed algorithm. Future research will focus on extending this framework to accommodate low-rank identification of nonlinear dynamical systems, addressing the current limitation of neglecting system nonlinearities.

\counterwithin*{lemma}{section} 
\renewcommand{\thelemma}{A\arabic{lemma}}

\counterwithin*{remark}{section}
\renewcommand{\theremark}{A\arabic{remark}}

\counterwithin*{definition}{section}
\renewcommand{\thedefinition}{A\arabic{definition}}

\appendices


\subsection{Preliminary Results}\label{app:pre}


\begin{lemma}(\cite{chen2013reduced})\label{lem:closedsol}
For any $\lambda \geq 0$, $w=[w_1,\cdots,w_n]^{\top}$ with $0 \leq w_1 \leq \cdots \leq w_{n}$, and $Y \in \mathbb{R}^{d \times n}$ with a singular value decomposition $Y=U \Sigma V^\top$, a global optimal solution to the optimization problem
\begin{align}
    \min_{X\in \mathbb{R}^{d\times n}} \frac{1}{2}\|Y-X\|^2+\lambda\|X\|_{w,*}
\end{align}
is $\mathcal{S}_{\lambda w}(Y)$, where
\begin{align}
    &\mathcal{S}_{\lambda w}(Y)=U \mathcal{S}_{\lambda w}(\Sigma) V^\top,\nonumber\\
    &\mathcal{S}_{\lambda w}(\Sigma)=
    \left[
    \begin{matrix}
        {\rm diag}\left\{\max\left\{\sigma_i(Y)-\lambda w_i, 0\right\}, i=1, \cdots, n\right\}\\
        0_{(d-n)\times n}
    \end{matrix}
    \right].\nonumber
\end{align}
Further, if all the nonzero singular values of $Y$ are distinct, $\mathcal{S}_{\lambda w}(Y)$ is the unique optimal solution. $\mathcal{S}_{\lambda w}(Y)$ is also referred to as the soft singular value thresholding operator.
\end{lemma}


\begin{lemma}(\cite{matrixanalysis2012})\label{lem:singularvalue}
For any two matrices $A,~B\in \mathbb{R}^{d\times n}$, it holds that
\begin{align}
\sum_{i=1}^{\min\{d,n\}} \left(\sigma_i(A)-\sigma_i(B)\right)^2\leq \left\|A-B\right\|^2,
\end{align}
where the singular values of matrices are arranged in a descending order.
\end{lemma}

The following result establishes the asymptotical error bounds of the LS algorithm (\cite{chen2012identification}).


\begin{lemma}(\cite{chen2012identification})\label{lem:lserr}
Assume that Assumptions \ref{assum:1} and \ref{assum:2} hold. Then
\begin{align}
    \| \Theta_{N+1}-\Theta \|=O \left(\sqrt{\frac{\log\lambda_{\max}(N)}{\lambda_{\min}(N)}}\right),~~\mathrm{a.s.}
\end{align}
\end{lemma}

\begin{lemma}(Corollary 3 of Section 9.5 in \cite{chow2012probability})\label{lem:clt}
For each $N\geq 1$, let $\{S_{N,l}=\sum_{k=1}^l \zeta_{N,k},~1\leq l\leq N<\infty\}$ be an $\mathcal{L}_2$ random sequence with index $l$ on  $\left(\Omega,\mathcal{F},\mathbb{P}\right)$ satisfying
\begin{align}
\sum_{k=1}^N &\mathbb{E}\left[\left.\zeta_{N,k+1}\right|\mathcal{F}_k\right]
\mathop{\longrightarrow} \limits_{N \rightarrow \infty}^{\mathbb{P}} 0, ~
\sum_{k=1}^N \left(\mathbb{E}\left[\left.\zeta_{N,k+1}\right|\mathcal{F}_k\right]\right)^2
\mathop{\longrightarrow} \limits_{N \rightarrow \infty}^{\mathbb{P}} 0\\
\sum_{k=1}^N &\left( \mathbb{E}\left[\left.\zeta_{N,k+1}^2\right|\mathcal{F}_k\right]-\left(\mathbb{E}\left[\left.\zeta_{N,k+1}\right|\mathcal{F}_k\right]\right)^2\right)
\mathop{\longrightarrow} \limits_{N \rightarrow \infty}^{\mathbb{P}} \sigma^2, \\
\sum_{k=1}^N &\mathbb{E}\left[\left.\zeta_{N,k+1}^2 \mathbb{I}\left\{\left|\zeta_{N,k+1}\right|>\varepsilon\right\}\right|\mathcal{F}_k\right]
\mathop{\longrightarrow} \limits_{N \rightarrow \infty}^{\mathbb{P}} 0,~ \forall \epsilon>0
\end{align}
for some constant $\sigma>0$. Then it follows that
\begin{align}
S_{N,N} \mathop{\longrightarrow} \limits_{N \rightarrow \infty}^{\mathcal{D}} \mathcal{N}\left(0,\sigma^2\right).
\end{align}
\end{lemma}

\begin{lemma}(Cram\'{e}r-Wold Device (Theorem 8.4 in \cite{hansen2022probability}))\label{lem:cw}
For random vectors $x_N\in \mathbb{R}^n,~N\geq1$ and $x\in \mathbb{R}^n$, $x_N \mathop{\longrightarrow} \limits_{N \rightarrow \infty}^\mathcal{D} x$ if and only if $v^\top x_N \mathop{\longrightarrow} \limits_{N \rightarrow \infty}^\mathcal{D} v^\top x$, for any deterministic vector $v\in \mathbb{R}^n$.
\end{lemma}


\subsection{Proof of Lemma \ref{lem:JN}}\label{app:lem1}


Notably the weights of the optimization problem (\ref{pro:J}) are $w_{N+1}(i)=\left(\widehat{\sigma}_{i}\left(\Theta_{N+1}\right)\right)^{-1}$, $i\in \left\{1,\cdots,n\right\}$ and for the singular values of $\Theta_{N+1}$, it holds that $\sigma_1(\Theta_{N+1})\geq\cdots\geq\sigma_n(\Theta_{N+1})$. This indicates $0\leq w_{N+1}(1)\leq \cdots \leq w_{N+1}(n)$. By Lemma \ref{lem:closedsol}, it follows that an optimal solution to (\ref{pro:J}) is given as follows:
\begin{align}\label{eq:s1}
X_{N+1}\!\!
&\triangleq \!\!
U_{N+1}
\left[
    \begin{smallmatrix}
        {\rm diag}\left\{\max\left\{\sigma_i(\Theta_{N+1})-\lambda_N w_{N+1}(i), 0\right\}, i=1, \cdots, n\right\}\\
        0_{(d-n)\times n}
    \end{smallmatrix}
    \right]
V_{N+1}^\top\nonumber\\
&\in \mathop{{\rm argmin}} \limits_{X \in \mathbb{R}^{d \times n}} J_{N+1}(X).
\end{align}
This finishes the proof.

\subsection{Proof of Theorem \ref{th:rank}}\label{app:thrank}

Noting that Assumption \ref{assum:3} holds almost surely, there exists an $\omega$-set $\Omega_0$ with $\mathbb{P}\{\Omega_0\}=1$ such that Assumption \ref{assum:3} holds for any $\omega \in \Omega_0$. In the subsequent analysis, we will consider a fixed sample path $\omega \in \Omega_0$.

Recall the definition of the rank estimate $\widehat{r}_N$,
\begin{align}
\widehat{r}_N\triangleq \max \left\{i\left|1\leq j \leq i,~\sigma_j\left(\Theta_{N+1}\right)\geq\lambda_N w_{N+1}(j)\right.\right\}.\nonumber
\end{align}

Since $\mathrm{rank}(\Theta)=r$ and $\sigma_i(\Theta),~i=1,\cdots,n$ are the singular values of $\Theta$ arranged in a descending order, for any $i\in\{1,\cdots,r\}$, we have $\sigma_i(\Theta)\neq 0$.

By the definitions of $w_{N+1}(i)$ and $\widehat{\sigma}_i\left(\Theta_{N+1}\right)$, it directly follows that
\begin{align}\label{eq:rcons1}
\sigma_i\left(\Theta_{N+1}\right)-\lambda_N w_{N+1}(i)
&=\frac{\sigma_i\left(\Theta_{N+1}\right)\widehat{\sigma}_i\left(\Theta_{N+1}\right)-\lambda_N}{\widehat{\sigma}_i\left(\Theta_{N+1}\right)}
\end{align}
and
\begin{align}\label{eq:rcons2}
&\sigma_i\left(\Theta_{N+1}\right)\widehat{\sigma}_i\left(\Theta_{N+1}\right)-\lambda_N\nonumber\\
&=\sigma_i\left(\Theta_{N+1}\right)\left({\sigma}_i\left(\Theta_{N+1}\right)+\sqrt{\frac{\log\lambda_{\max}(N)}{\lambda_{\min}(N)}}\right)-\lambda_N\nonumber\\
&=\left(\sigma_i\left(\Theta_{N+1}\right)\right)^2+{\sigma}_i\left(\Theta_{N+1}\right)\sqrt{\frac{\log\lambda_{\max}(N)}{\lambda_{\min}(N)}}-\lambda_N.
\end{align}

By Lemmas \ref{lem:singularvalue} and \ref{lem:lserr} and Assumption \ref{assum:3}, for any $i\in\{1,\cdots,n\}$, we have
\begin{align}
&\left|\sigma_i\left(\Theta_{N+1}\right)-\sigma_i(\Theta)\right|\leq \sqrt{\sum_{i=1}^n \left( \sigma_i\left(\Theta_{N+1}\right)-\sigma_i(\Theta)\right)^2}\nonumber\\
&\leq \left\|\Theta_{N+1}-\Theta\right\|
=O\left(\sqrt{\frac{\log\lambda_{\max}(N)}{\lambda_{\min}(N)}}\right)\mathop{\longrightarrow}\limits_{N\to\infty}0.\label{eq20}
\end{align}

If $i\in\{1,\cdots,r\}$, since $\sigma_i\left(\Theta_{N+1}\right)\mathop{\longrightarrow} \limits_{N \rightarrow \infty} \sigma_i(\Theta) \neq 0$ and noting Assumption \ref{assum:lambdan01} that $\lambda_N \mathop{\longrightarrow} \limits_{N \rightarrow \infty} 0$, there exists an integer $N_1^i>0$ large enough such that for all $N\geq N_1^i$, $\sigma_i\left(\Theta_{N+1}\right)\widehat{\sigma}_i\left(\Theta_{N+1}\right)-\lambda_N> 0$ which in turn guarantees $\sigma_i\left(\Theta_{N+1}\right)-\lambda_N w_{N+1}(i)>0$ by using (\ref{eq:rcons1}). Then from the definition of $\widehat{r}_N$ we have that $r \leq \widehat{r}_N,~\forall N\geq N_1^i$.

If $i\in\{r+1,\cdots,n\}$, since $\mathrm{rank}(\Theta)=r$ we have $\sigma_i(\Theta)=0$ and by (\ref{eq20}), $\sigma_i\left(\Theta_{N+1}\right)=O\left(\sqrt{\frac{\log\lambda_{\max}(N)}{\lambda_{\min}(N)}}\right)$ and $\widehat{\sigma}_i\left(\Theta_{N+1}\right)=O\left(\sqrt{\frac{\log\lambda_{\max}(N)}{\lambda_{\min}(N)}}\right)$. Then by Assumption \ref{assum:lambdan01}, there exists a constant $c>0$ such that
\begin{align}
\frac{\lambda_N w_{N+1}(i)}{\sigma_i\left(\Theta_{N+1}\right)}\geq c \frac{\lambda_N}{\frac{\log\lambda_{\max}(N)}{\lambda_{\min}(N)}} \mathop{\longrightarrow} \limits_{N \rightarrow \infty} \infty,
\end{align}
which leads to the fact that for some positive integer $N_2^i>0$ large enough such that for any $N\geq N_2^i$, $\sigma_i\left(\Theta_{N+1}\right)<\lambda_N w_{N+1}(i)$. Then from the definition of $\widehat{r}_N$ we have that $\widehat{r}_N\leq r,~\forall N\geq N_2^i$.

Setting $N_0\triangleq \max\{N_1^i,~i\in\{1,\cdots,r\},~N_2^j,~j\in\{r+1,\cdots,n\}\}$, it holds that $\widehat{r}_N=r,~\forall N\geq N_0$. This proves (\ref{eq11}). Based on the above analysis and (\ref{eq:defXN}), it follows that (\ref{eq12}) holds. This finishes the proof.

\subsection{Proof of Theorem \ref{th:errorbound}}\label{app:therrorbound}

By Theorem \ref{th:rank}, there exists an $\omega$-set $\Omega_0$ with $\mathbb{P}\{\Omega_0\}=1$ such that for any $\omega \in \Omega_0$, there exists an integer $N_0(\omega)>0$ such that
\begin{align}\label{eq:rtheta}
\widehat{r}_N={\rm rank}(X_{N+1})=r,~N \geq N_0(\omega)
\end{align}
and
\begin{align}\label{eq:defXN1}
    &X_{N+1}=U_{N+1}\nonumber\\
    & \cdot
    \left[
    \begin{smallmatrix}
        {\rm diag}\left\{\sigma_i(\Theta_{N+1})-\lambda_N w_{N+1}(i), i=1,\cdots,r\right\} & 0_{r\times (n-r)}\\
        0_{(d-r)\times r} & 0_{(d-r)\times (n-r)}
    \end{smallmatrix}
    \right]
    V_{N+1}^\top.
\end{align}

Note that for any matrix $A\in \mathbb{R}^{d\times n}$, the equality holds for its Frobenius norm, i.e., $\|A\|=\sqrt{{\rm tr}\left(A^\top A\right)}=\sqrt{\sum_{i=1}^{\min\{d,n\}} \sigma_i^2(A)}$. From (\ref{eq:defXN1}) we can get
\begin{align}\label{eq:xnerror}
    &\|X_{N+1}-\Theta\|\nonumber\\
    &\leq \left\|X_{N+1}-\Theta_{N+1}\right\|+\left\|\Theta_{N+1}-\Theta\right\|\nonumber\\
    &= \sqrt{\sum_{i=1}^n \left(\sigma_{i}\left(X_{N+1}\right)-\sigma_{i}\left(\Theta_{N+1}\right)\right)^2}+\left\|\Theta_{N+1}-\Theta\right\|\nonumber\\
    &= \sqrt{\sum_{i=1}^{r} \left(\lambda_N w_{N+1}(i)\right)^2+\sum_{i=r+1}^n \left(\sigma_{i}\left(\Theta_{N+1}\right)\right)^2}\nonumber\\
    &~~~+\left\|\Theta_{N+1}-\Theta\right\|.
\end{align}

For any $i\in\{1,\cdots,n\}$, by Lemmas \ref{lem:singularvalue} and \ref{lem:lserr}, for the LS estimates it holds that
\begin{align}
    \left|\sigma_i(\Theta_{N+1})-\sigma_i(\Theta)\right|
    &\leq \sqrt{\sum_{i=1}^n \left(\sigma_{i}\left(\Theta_{N+1}\right)-\sigma_i(\Theta)\right)^2}\nonumber\\
    &\leq\left\|\Theta_{N+1}-\Theta\right\|=O\left(\sqrt{\frac{\log \lambda_{\max}(N)}{\lambda_{\min}(N)}}\right).
\end{align}

For any $i\in\{1,\cdots,r\}$, since $\sigma_i(\Theta)>0$ it yields that there exist constants $c>0$ and $C>0$ such that
\begin{align}
    \sigma_i(\Theta_{N+1})
    &\geq \sigma_i(\Theta)-\left|\sigma_i(\Theta_{N+1})-\sigma_i(\Theta)\right|\nonumber\\
    &\geq \sigma_i(\Theta)-c\sqrt{\frac{\log \lambda_{\max}(N)}{\lambda_{\min}(N)}}
    >C
\end{align}
and
\begin{align}
    \frac{1}{\widehat{\sigma}_i(\Theta_{N+1})}
    =\frac{1}{{\sigma}_i\left(\Theta_{N+1}\right)+\sqrt{\frac{\log\lambda_{\max}(N)}{\lambda_{\min}(N)}}}
    \leq \frac{1}{{\sigma}_i\left(\Theta_{N+1}\right)}
    <\frac{1}{C}.
\end{align}
Then we have
\begin{align}\label{eq:sig2thetan0}
    \sum_{i=1}^{r} \left(\lambda_N w_{N+1}(i)\right)^2
    =\sum_{i=1}^{r} \left(\frac{\lambda_N}{\widehat{\sigma}_i(\Theta_{N+1})}\right)^2
    =O\left(\lambda^2_N\right).
\end{align}

For any $i\in\{r+1,\cdots,n\}$, since $\sigma_i(\Theta)=0$ it yields that
\begin{align}\label{eq:sig2thetan}
    \sum_{i=r+1}^n \left(\sigma_{i}\left(\Theta_{N+1}\right)\right)^2
    &=\sum_{i=r+1}^n \left(\sigma_i\left(\Theta_{N+1}\right)-\sigma_i(\Theta)\right)^2\nonumber\\
    &\leq\sum_{i=1}^n \left(\sigma_i\left(\Theta_{N+1}\right)-\sigma_i(\Theta)\right)^2\nonumber\\
    &\leq \left\|\Theta_{N+1}-\Theta\right\|^2
    =O\left(\frac{\log \lambda_{\max}(N)}{\lambda_{\min}(N)}\right).
\end{align}

Substituting (\ref{eq:sig2thetan0}) and (\ref{eq:sig2thetan}) into (\ref{eq:xnerror}) and noting Assumption \ref{assum:lambdan01}, we obtain
\begin{align}
&\|X_{N+1}-\Theta\|\nonumber\\
&=O\left(\sqrt{\lambda^2_N+\frac{\log \lambda_{\max}(N)}{\lambda_{\min}(N)}}\right)+O\left(\sqrt{\frac{\log \lambda_{\max}(N)}{\lambda_{\min}(N)}}\right)\nonumber\\
&=O\left(\sqrt{\frac{\log \lambda_{\max}(N)}{\lambda_{\min}(N)}}\right).\label{eq32}
\end{align}

From (\ref{eq32}) and by Assumption \ref{assum:3}, it directly follows that
\begin{align}
X_{N+1}(s,t) \mathop{\longrightarrow} \limits_{N \rightarrow \infty} \Theta(s,t),~ s=1,\cdots,d,~t=1,\cdots,n,~~ \mathrm{a.s.}
\end{align}
This finishes the proof.

\subsection{Proof of Lemma \ref{lem:u1u1t}}\label{app:lemu1u1t}

\begin{sloppypar}
Recall that the SVD of $\Theta_{N+1}$ and $\Theta$ are $\Theta_{N+1}=U_{N+1} \Sigma_{N+1} V_{N+1}^\top$ and $\Theta=U \Sigma V^\top$, respectively, where ${\Sigma}_{N+1} = \left[{\rm diag}\left\{\sigma_{1}\left(\Theta_{N+1}\right),\cdots,\sigma_{n}\left(\Theta_{N+1}\right)\right\};0_{(d-n)\times n}\right]$ and ${\Sigma} = \left[{\rm diag}\left\{\sigma_{1}\left(\Theta\right),\cdots,\sigma_{n}\left(\Theta\right)\right\};0_{(d-n)\times n}\right]$.
\end{sloppypar}

For simplicity of notations, denote $\sigma_{N+1,i}\triangleq \sigma_{i} \left(\Theta_{N+1}\right)$ and $\sigma_{i}\triangleq \sigma_{i} \left(\Theta\right)$ for any $i\in \{1,\cdots,n\}$. It follows that
\begin{align}
&\Theta_{N+1} \Theta_{N+1}^\top \nonumber\\
&=U_{N+1} {\rm diag}\Big\{\sigma_{N+1,1}^2,\cdots,\sigma_{N+1,n}^2,\underbrace{0,\cdots,0}_{d-n}\Big\} U_{N+1}^\top,\label{eq42}
\end{align}
and
\begin{align}
\Theta \Theta^\top =U {\rm diag}\Big\{\sigma_{1}^2,\cdots,\sigma_{n}^2,\underbrace{0,\cdots,0}_{d-n}\Big\} U^\top. \label{eq43}
\end{align}

From (\ref{eq42}) and (\ref{eq43}), the $(s,t)$-component of $\Theta_{N+1} \Theta_{N+1}^\top$ and $\Theta \Theta^\top$ can be respectively represented as
\begin{align}
&\left[\Theta_{N+1} \Theta_{N+1}^\top\right] (s,t)
=\sum_{i=1}^{n} U_{N+1}(s,i) \sigma_{N+1,i}^2 U_{N+1}^\top(i,t)\nonumber\\
&=\sum_{i=1}^{n} \sigma_{N+1,i}^2 U_{N+1}(s,i) U_{N+1}(t,i),\label{eq44}
\end{align}
and
\begin{align}
\left[\Theta \Theta^\top\right] (s,t)=\sum_{i=1}^{n} \sigma_{i}^2 U(s,i) U(t,i),\label{eq45}
\end{align}
from which it follows that
\begin{align}
&\left[\Theta_{N+1} \Theta_{N+1}^\top\right] (s,t)-\left[\Theta \Theta^\top\right] (s,t)\nonumber\\
&=\sum_{i=1}^{n} \sigma_{N+1,i}^2 U_{N+1}(s,i) U_{N+1}(t,i)-\sum_{i=1}^{n} \sigma_{i}^2 U(s,i) U(t,i)\nonumber\\
&=\sum_{i=1}^{n} \left(\sigma_{N+1,i}^2-\sigma_{i}^2\right) U_{N+1}(s,i) U_{N+1}(t,i)\nonumber\\
&~~~+\sum_{i=1}^{n} \sigma_{i}^2 \left(U_{N+1}(s,i) U_{N+1}(t,i)-U(s,i) U(t,i)\right).\label{eq46}
\end{align}

From (\ref{eq46}), we have
\begin{align}\label{eq:item0}
&\sum_{i=1}^{n} \sigma_{i}^2 \left(U_{N+1}(s,i) U_{N+1}(t,i)-U(s,i) U(t,i)\right)\nonumber\\
&=\left[\Theta_{N+1} \Theta_{N+1}^\top\right] (s,t)-\left[\Theta \Theta^\top\right] (s,t)\nonumber\\
&~~~-\sum_{i=1}^{n} \left(\sigma_{N+1,i}^2-\sigma_{i}^2\right) U_{N+1}(s,i) U_{N+1}(t,i).
\end{align}

By the orthogonality of $U_{N+1}$, it yields that for any $s \in \{1,\cdots,d\}$,
\begin{align}
\sum_{i=1}^{d} U_{N+1}(s,i) U_{N+1}^\top(i,s)
&=\sum_{i=1}^{d} U_{N+1}(s,i) U_{N+1}(s,i)\nonumber\\
&=\sum_{i=1}^{d} U_{N+1}^2(s,i)=1
\end{align}
and thus for any $s \in \{1,\cdots,d\}$ and $i \in \{1,\cdots,d\}$
\begin{align}\label{eq:u}
\left|U_{N+1}(s,i) U_{N+1}(t,i)\right|
\leq \left|U_{N+1}(s,i)\right| \left|U_{N+1}(t,i)\right|\leq 1
\end{align}
which means that $\left\{U_{N+1}(s,i) U_{N+1}(t,i)\right\}_{N\geq 1}$ is bounded.

By Assumption \ref{assum:3} and Lemmas \ref{lem:singularvalue} and \ref{lem:lserr}, when $N\rightarrow \infty$, we have
\begin{align}
\Theta_{N+1} \mathop{\longrightarrow} \limits_{N \rightarrow \infty} \Theta,~\sigma_{N+1,i} \mathop{\longrightarrow} \limits_{N \rightarrow \infty} \sigma_{i},~~{\rm a.s.}
\end{align}
for $i\in \{1,\cdots,n\}$. Hence it follows that
\begin{align}\label{eq:item01}
&\left\|\Theta_{N+1} \Theta_{N+1}^\top-\Theta \Theta^\top\right\|\nonumber\\
&=\big\|\left(\Theta_{N+1}-\Theta\right)\left(\Theta_{N+1}^\top+\Theta^\top\right)+\Theta \left(\Theta_{N+1}^\top - \Theta^\top\right) \nonumber\\
&~~~+ \left(\Theta -\Theta_{N+1}\right) \Theta^\top\big\|\nonumber\\
&=O\left(\left\|\Theta_{N+1}-\Theta\right\|\right)
=O\left(\sqrt{\frac{\log \lambda_{\max}(N)}{\lambda_{\min}(N)}}\right)
\end{align}
and for $i\in \{1,\cdots,n\}$
\begin{align}\label{eq:0s}
&\left|\sigma_{N+1,i}^2-\sigma_{i}^2\right|
=\left|\left(\sigma_{N+1,i}-\sigma_{i}\right)\left(\sigma_{N+1,i}+\sigma_{i}\right)\right|\nonumber\\
&=O\left(\left|\sigma_{N+1,i}-\sigma_{i}\right|\right)=O\left(\sqrt{\frac{\log \lambda_{\max}(N)}{\lambda_{\min}(N)}}\right).
\end{align}

For any $j\in \{1,\cdots,n\}$, we consider
\begin{align}
&\prod_{l=1}^{j} \Theta_{N+1} \Theta_{N+1}^\top\nonumber\\
&=U_{N+1} {\rm diag}\Big\{\sigma_{N+1,1}^{2j},\cdots,\sigma_{N+1,n}^{2j},\underbrace{0,\cdots,0}_{d-n}\Big\} U_{N+1}^\top,\\
&\prod_{l=1}^{j} \Theta \Theta^\top
=U {\rm diag}\Big\{\sigma_{1}^{2j},\cdots,\sigma_{n}^{2j},\underbrace{0,\cdots,0}_{d-n}\Big\} U^\top.
\end{align}

Similar to (\ref{eq:item0}), we can obtain that for any $j\in \{1,\cdots,n\}$,
\begin{align}\label{eq:item}
&\sum_{i=1}^{n} \sigma_{i}^{2j} \left(U_{N+1}(s,i) U_{N+1}(t,i)-U(s,i) U(t,i)\right)\nonumber\\
&=\left[\prod_{l=1}^{j} \Theta_{N+1} \Theta_{N+1}^\top\right] (s,t)-\left[\prod_{l=1}^{j} \Theta \Theta^\top\right] (s,t)\nonumber\\
&~~~-\sum_{i=1}^{n} \left(\sigma_{N+1,i}^{2j}-\sigma_{i}^{2j}\right) U_{N+1}(s,i) U_{N+1}(t,i).
\end{align}

Carrying out a similar analysis as (\ref{eq:item01}) and (\ref{eq:0s}), we can prove that
\begin{align}
&\left\|\prod_{l=1}^{j} \Theta_{N+1} \Theta_{N+1}^\top-\prod_{l=1}^{j} \Theta \Theta^\top\right\|=O\left(\sqrt{\frac{\log \lambda_{\max}(N)}{\lambda_{\min}(N)}}\right),\label{eq:item1}\\
&\left|\sigma_{N+1,i}^{2j}-\sigma_{i}^{2j}\right|=O\left(\sqrt{\frac{\log \lambda_{\max}(N)}{\lambda_{\min}(N)}}\right),\label{eq:s}
\end{align}
for any $j\in\{1,\cdots,n\}$ and $i\in\{1,\cdots,n\}$.

Combining (\ref{eq:u}) and (\ref{eq:s}), for $i\in \{1,\cdots,n\}$ and $j\in \{1,\cdots,n\}$, we have
\begin{align}\label{eq:item2}
&\left|\left(\sigma_{N+1,i}^{2j}-\sigma_{i}^{2j}\right)U_{N+1}(s,i) U_{N+1}(t,i)\right|\nonumber\\
&=O\left(\sqrt{\frac{\log \lambda_{\max}(N)}{\lambda_{\min}(N)}}\right).
\end{align}

Noting that $\sigma_{i}=0$ for any $i\in \{r+1,\cdots,n\}$, we can obtain
\begin{align}\label{eq:sigr}
&\left[
\begin{smallmatrix}
\sigma_1^2 & \sigma_2^2 & \cdots & \sigma_r^2 \\
\vdots & \vdots & & \vdots \\
\sigma_1^{2r} & \sigma_2^{2r} & \cdots & \sigma_r^{2r}
\end{smallmatrix}
\right]
\left[
\begin{smallmatrix}
U_{N+1}(s,1) U_{N+1}(t,1)-U(s,1) U(t,1)\\
\vdots \\
U_{N+1}(s,r) U_{N+1}(t,r)-U(s,r) U(t,r)
\end{smallmatrix}
\right]\nonumber\\
&=\left[
\begin{smallmatrix}
\sigma_1^2 & \sigma_2^2 & \cdots & \sigma_n^2 \\
\vdots & \vdots & & \vdots \\
\sigma_1^{2r} & \sigma_2^{2r} & \cdots & \sigma_n^{2r}
\end{smallmatrix}
\right]
\left[
\begin{smallmatrix}
U_{N+1}(s,1) U_{N+1}(t,1)-U(s,1) U(t,1)\\
\vdots \\
U_{N+1}(s,n) U_{N+1}(t,n)-U(s,n) U(t,n)
\end{smallmatrix}
\right]\nonumber
\end{align}
\begin{align}
&=\left[
\begin{smallmatrix}
\left[\prod_{l=1}^{1} \Theta_{N+1} \Theta_{N+1}^\top\right](s,t)-\left[\prod_{l=1}^{1} \Theta \Theta^\top\right](s,t)\\
\vdots\\
\left[\prod_{l=1}^{r} \Theta_{N+1} \Theta_{N+1}^\top\right](s,t)-\left[\prod_{l=1}^{r} \Theta \Theta^\top\right](s,t)
\end{smallmatrix}
\right]\nonumber\\
&~~~-
\left[
\begin{smallmatrix}
\sum_{i=1}^{n} \left(\sigma_{N+1,i}^2-\sigma_{i}^2\right) U_{N+1}(s,i) U_{N+1}(t,i)\\
\vdots\\
\sum_{i=1}^{n} \left(\sigma_{N+1,i}^{2r}-\sigma_{i}^{2r}\right) U_{N+1}(s,i) U_{N+1}(t,i)
\end{smallmatrix}
\right]
\end{align}
where the last equality holds by (\ref{eq:item}) for $j \in \{1,\cdots,r\}$.

Since by Assumption \ref{assum:theta} the nonzero singular values of $\Theta$ are pairwise distinct, the matrix
$\left[
\begin{smallmatrix}
\sigma_1^2 & \sigma_2^2 & \cdots & \sigma_r^2 \\
\vdots & \vdots & & \vdots \\
\sigma_1^{2r} & \sigma_2^{2r} & \cdots & \sigma_r^{2r}
\end{smallmatrix}
\right]$
is nonsingular. 

Multiplying both sides of (\ref{eq:sigr}) by the inverse matrix of
$\left[
\begin{smallmatrix}
\sigma_1^2 & \sigma_2^2 & \cdots & \sigma_r^2 \\
\vdots & \vdots & & \vdots \\
\sigma_1^{2r} & \sigma_2^{2r} & \cdots & \sigma_r^{2r}
\end{smallmatrix}
\right]$, we can obtain that
\begin{align}
&\left[
        \begin{smallmatrix}
            U_{N+1}(s,1) U_{N+1}(t,1)-U(s,1) U(t,1)\\
            \vdots \\
            U_{N+1}(s,r) U_{N+1}(t,r)-U(s,r) U(t,r)
        \end{smallmatrix}
\right]\nonumber\\
&=\left[
        \begin{smallmatrix}
            \sigma_1^2 & \sigma_2^2 & \cdots & \sigma_r^2 \\
            \vdots & \vdots & & \vdots \\
            \sigma_1^{2r} & \sigma_2^{2r} & \cdots & \sigma_r^{2r}
        \end{smallmatrix}
\right]^{-1}
\left[
        \begin{smallmatrix}
            \left[\prod_{l=1}^{1} \Theta_{N+1} \Theta_{N+1}^\top\right](s,t)-\left[\prod_{l=1}^{1} \Theta \Theta^\top\right](s,t)\\
            \vdots\\
            \left[\prod_{l=1}^{r} \Theta_{N+1} \Theta_{N+1}^\top\right](s,t)-\left[\prod_{l=1}^{r} \Theta \Theta^\top\right](s,t)
        \end{smallmatrix}
\right]\nonumber\\
        &~~~-
        \left[
        \begin{smallmatrix}
            \sigma_1^2 & \sigma_2^2 & \cdots & \sigma_r^2 \\
            \vdots & \vdots & & \vdots \\
            \sigma_1^{2r} & \sigma_2^{2r} & \cdots & \sigma_r^{2r}
        \end{smallmatrix}
\right]^{-1}
\left[
        \begin{smallmatrix}
            \sum_{i=1}^{n} \left(\sigma_{N+1,i}^2-\sigma_{i}^2\right) U_{N+1}(s,i) U_{N+1}(t,i)\\
            \vdots\\
            \sum_{i=1}^{n} \left(\sigma_{N+1,i}^{2r}-\sigma_{i}^{2r}\right) U_{N+1}(s,i) U_{N+1}(t,i)
        \end{smallmatrix}
\right].\label{eq64}
\end{align}

From (\ref{eq64}) and by noting (\ref{eq:item1}) and (\ref{eq:s}), it yields that
\begin{align}
\left\|
\left[
        \begin{smallmatrix}
            U_{N+1}(s,1) U_{N+1}(t,1)-U(s,1) U(t,1)\\
            \vdots \\
            U_{N+1}(s,r) U_{N+1}(t,r)-U(s,r) U(t,r)
        \end{smallmatrix}
\right]
\right\|=O\left(\sqrt{\frac{\log \lambda_{\max}(N)}{\lambda_{\min}(N)}}\right)\nonumber
\end{align}
and
\begin{align}\label{eq:element}
&\left|U_{N+1}(s,j) U_{N+1}(t,j)-U(s,j) U(t,j)\right|\nonumber\\
&=O\left(\sqrt{\frac{\log \lambda_{\max}(N)}{\lambda_{\min}(N)}}\right)
\end{align}
for $s\in \{1,\cdots,d\}$, $t\in \{1,\cdots,d\}$, and $j\in \{1,\cdots,r\}$.

On the other hand, 
the $(s,t)$-component of $U_{N+1}^{(1)} {U_{N+1}^{(1)}}^\top-U^{(1)} {U^{(1)}}^\top$ is
\begin{align}
&\left[U_{N+1}^{(1)} {U_{N+1}^{(1)}}^\top\right](s,t)-\left[U^{(1)} {U^{(1)}}^\top\right](s,t)\nonumber\\
&=\sum_{j=1}^r \Big(U_{N+1}(s,j) U_{N+1}(t,j)-U(s,j) U(t,j)\Big)
\end{align}
from which and by (\ref{eq:element}),
\begin{align}
&\left\|U_{N+1}^{(1)} {U_{N+1}^{(1)}}^\top - U^{(1)} {U^{(1)}}^\top\right\|\nonumber\\
&\leq \sum_{s=1}^d \sum_{t=1}^d \left|\sum_{j=1}^r \Big(U_{N+1}(s,j) U_{N+1}(t,j)-U(s,j) U(t,j)\Big)\right|\nonumber\\
&=O\left(\sqrt{\frac{\log \lambda_{\max}(N)}{\lambda_{\min}(N)}}\right).
\end{align}
This finishes the proof.

\subsection{Proof of Lemma \ref{lem:LSasy}}\label{app:lemLSasy}

By the properties of matrix Kronecker product, direct calculation leads to
\begin{align}\label{eq:zasy3}
&{\rm vec}\left(\left(\sum_{k=1}^N \varphi_k \varphi_k^\top\right)^{-1/2} \sum_{k=1}^N \varphi_k \varepsilon_{k+1}^\top\right)\nonumber\\
&=\left( I_n \otimes \left(\sum_{k=1}^N \varphi_k \varphi_k^\top\right)^{-1/2}\right) \sum_{k=1}^N \left(I_n \otimes \varphi_k\right) \varepsilon_{k+1}.
\end{align}

By Assumption \ref{assum:5}, we have
\begin{align}\label{eq:zasy2}
    &\left( I_n \otimes \left(\sum_{k=1}^N \varphi_k \varphi_k^\top\right)^{-1/2}\right) \left(I_n \otimes B_N\right)\nonumber\\
    &=I_n \otimes \left(\left(\sum_{k=1}^N \varphi_k \varphi_k^\top\right)^{-1/2} B_N\right)
    \mathop{\longrightarrow} \limits_{N \rightarrow \infty}^{\mathbb{P}} I_{dn}.
\end{align}

By (\ref{eq:zasy3}) and (\ref{eq:zasy2}), for (\ref{eq19}) we just need to prove
\begin{align}\label{eq:zasy1}
    \left( I_n \otimes B_N\right)^{-1} \sum_{k=1}^N \left(I_n \otimes \varphi_k\right) \varepsilon_{k+1}
    \mathop{\longrightarrow} \limits_{N \rightarrow \infty}^\mathcal{D}
    \mathcal{N}\left(0_{dn}, \sigma^2 I_{dn}\right).
\end{align}

By Lemma \ref{lem:cw}, for (\ref{eq:zasy1}) it suffices to verify that for any deterministic vector $v\in \mathbb{R}^{dn}$,
\begin{align}\label{eq:zasy}
    v^\top \left( I_n \otimes B_N\right)^{-1} \sum_{k=1}^N \left(I_n \otimes \varphi_k\right) \varepsilon_{k+1}
    \mathop{\longrightarrow} \limits_{N \rightarrow \infty}^\mathcal{D}
    \mathcal{N}\left(0, \sigma^2 \|v\|^2\right).
\end{align}

The subsequent proof is motivated by \cite{lai1981consistency}. The key difference lies in that here we consider the multivariate case and this makes the analysis procedure much involved. Here we present the details. 


For the fixed deterministic vector $v\in \mathbb{R}^{dn}$, denote
$$z_{N,k} \triangleq v^\top \left( I_n \otimes B_N\right)^{-1} \left(I_n \otimes \varphi_k\right),~k=1,\cdots,N$$
and
$$z_{N,k}^\prime \triangleq z_{N,k} \mathbb{I}\left\{\left\|z_{N,k}\right\|\leq 1\right\},~k=1,\cdots,N.$$

Obviously, $z_{N,k}$ and $z_{N,k}^\prime$ are $\mathcal{F}_k$-measurable and by Assumption \ref{assum:4}, it follows that
\begin{align}\label{eq:zp0}
    \mathbb{E}\left[\left.z_{N,k}^\prime \varepsilon_{k+1} \right| \mathcal{F}_k\right]
    =z_{N,k}^\prime \mathbb{E}\left[\left.\varepsilon_{k+1} \right| \mathcal{F}_k\right]
    =0,
\end{align}
and thus
\begin{align}\label{eq:condition1}
    \sum_{k=1}^N \mathbb{E}\left[\left.z_{N,k}^\prime \varepsilon_{k+1} \right|\mathcal{F}_k\right]
    =\sum_{k=1}^N \left(\mathbb{E}\left[\left.z_{N,k}^\prime \varepsilon_{k+1} \right|\mathcal{F}_k\right]\right)^2
    =0.
\end{align}

By Assumption \ref{assum:4} and the definition of $z_{N,k}^\prime$, it can be directly verified that $z_{N,k}^\prime \varepsilon_{k+1}$ is an $\mathcal{L}_2$ random sequence.

Noting that $( I_n \otimes B_N)^{-1} (I_n \otimes \varphi_k)=( I_n \otimes B_N^{-1}) (I_n \otimes \varphi_k)=I_n \otimes B_N^{-1} \varphi_k$ and Assumption \ref{assum:5}, we can further obtain
\begin{align}
    &\max_{1 \leq k \leq N}\left\|\left( I_n \otimes B_N\right)^{-1} \left(I_n \otimes \varphi_k\right)\right\|\nonumber\\
    &=\max_{1 \leq k \leq N}\left\| I_n \otimes B_N^{-1} \varphi_k\right\|
    \mathop{\longrightarrow} \limits_{N \rightarrow \infty}^{\mathbb{P}} 0
\end{align}
and thus
\begin{align}\label{eq:zpto0}
    \max_{1 \leq k \leq N}\left\|z_{N,k}\right\|
    &\leq \|v\| \cdot \max_{1 \leq k \leq N}\left\| \left( I_n \otimes B_N\right)^{-1} \left(I_n \otimes \varphi_k\right)\right\|\nonumber\\
    &\mathop{\longrightarrow} \limits_{N \rightarrow \infty}^{\mathbb{P}} 0.
\end{align}

Hence we know that
\begin{align}
    \mathbb{P}\left\{\max_{1 \leq k \leq N}\left\|z_{N,k}\right\|>1\right\}
    \mathop{\longrightarrow} \limits_{N \rightarrow \infty} 0,
\end{align}
and equivalently,
\begin{align}\label{eq:zzp}
    \mathbb{P}\left\{z_{N,k}\neq z_{N,k}^\prime, \text{ for some } 1\leq k \leq N\right\}
    \mathop{\longrightarrow} \limits_{N \rightarrow \infty} 0.
\end{align}

Next, by the definition of $z_{N,k}$ we have the following equalities,
\begin{align}\label{eq:zsq0}
    &\sum_{k=1}^N \mathbb{E}\left[\left.\left(z_{N,k} \varepsilon_{k+1}\right)^2\right|\mathcal{F}_k\right]\nonumber\\
    &=\sum_{k=1}^N \mathbb{E}\left[\left.z_{N,k} \varepsilon_{k+1} \varepsilon_{k+1}^\top z_{N,k}^\top\right|\mathcal{F}_k\right]\nonumber\\
    &=\sum_{k=1}^N \mathbb{E}\big[v^\top \left( I_n \otimes B_N^{-1}\right) \left(I_n \otimes \varphi_k\right) \varepsilon_{k+1} \varepsilon_{k+1}^\top \nonumber\\
    &~~~~~~~~~~~\cdot\left(I_n \otimes \varphi_k^\top\right) \left( I_n \otimes B_N^{-1}\right) v\big|\mathcal{F}_k\big].
\end{align}

Since the matrix $B_N$ is deterministic and $\varphi_k\in \mathcal{F}_k$, from (\ref{eq:zsq0}) we can further obtain
\begin{align}\label{eq:zsq}
    &\sum_{k=1}^N \mathbb{E}\left[\left.\left(z_{N,k} \varepsilon_{k+1}\right)^2\right|\mathcal{F}_k\right]\nonumber\\
    &=v^\top \left( I_n \otimes B_N^{-1}\right) \left(\sum_{k=1}^N \left(I_n \otimes \varphi_k\right) \mathbb{E}\left[\left. \varepsilon_{k+1} \varepsilon_{k+1}^\top\right|\mathcal{F}_k\right] \left(I_n \otimes \varphi_k^\top\right)\right) \nonumber\\
    &~~~\cdot \left( I_n \otimes B_N^{-1}\right) v\nonumber\\
    &=\sigma^2 v^\top \left( I_n \otimes B_N^{-1}\right) \left(\sum_{k=1}^N \left(I_n \otimes \varphi_k\right) \left(I_n \otimes \varphi_k^\top\right)\right) \left( I_n \otimes B_N^{-1}\right) v\nonumber\\
    &=\sigma^2 v^\top \left( I_n \otimes B_N^{-1} \left(\sum_{k=1}^N \varphi_k \varphi_k^\top\right)B_N^{-1}\right) v
    \mathop{\longrightarrow} \limits_{N \rightarrow \infty}^{\mathbb{P}} \sigma^2 \|v\|^2
\end{align}
where for the last limit taking place Assumption \ref{assum:5} is applied.

Combining (\ref{eq:condition1}), (\ref{eq:zzp}) and (\ref{eq:zsq}), it can be proved that
\begin{align}\label{eq:zpesq}
    &\sum_{k=1}^N \left(\mathbb{E}\left[\left.\left(z_{N,k}^\prime \varepsilon_{k+1}\right)^2\right|\mathcal{F}_k\right]-\left(\mathbb{E}\left[\left.z_{N,k}^\prime \varepsilon_{k+1} \right|\mathcal{F}_k\right]\right)^2\right)\nonumber\\
    &\mathop{\longrightarrow} \limits_{N \rightarrow \infty}^{\mathbb{P}} \sigma^2 \|v\|^2.
\end{align}

For any $\varepsilon>0$, by noting $z_{N,k}^\prime \in \mathcal{F}_k$, we have the following chain of inequality and equalities,
\begin{align}
    &\sum_{k=1}^N \mathbb{E}\left[\left.\left(z_{N,k}^\prime \varepsilon_{k+1}\right)^2 \mathbb{I}\left\{\left|z_{N,k}^\prime \varepsilon_{k+1}\right|>\varepsilon\right\}\right|\mathcal{F}_k\right]\nonumber\\
    &\leq \sum_{k=1}^N \mathbb{E}\left[\left.\left(z_{N,k}^\prime \varepsilon_{k+1}\right)^2 \cdot \frac{\left|z_{N,k}^\prime \varepsilon_{k+1}\right|^{\beta-2}}{\varepsilon^{\beta-2}}\right|\mathcal{F}_k\right] \nonumber\\
    &= \sum_{k=1}^N \mathbb{E}\left[\left.\left|z_{N,k}^\prime \varepsilon_{k+1}\right|^\beta \cdot \frac{1}{\varepsilon^{\beta-2}}\right|\mathcal{F}_k\right] \nonumber\\
    &\leq \sum_{k=1}^N \mathbb{E}\left[\left.\left\|z_{N,k}^\prime\right\|^\beta \cdot \left\|\varepsilon_{k+1}\right\|^\beta \cdot \frac{1}{\varepsilon^{\beta-2}}\right|\mathcal{F}_k\right] \nonumber\\
    &= \sum_{k=1}^N \frac{\left\|z_{N,k}^\prime\right\|^\beta}{\varepsilon^{\beta-2}}\cdot\mathbb{E}\left[\left.\left\|\varepsilon_{k+1}\right\|^\beta\right|\mathcal{F}_k\right] \nonumber\\
    &= \sum_{k=1}^N \mathbb{E}\left[\left.\left(z_{N,k}^\prime \varepsilon_{k+1}\right)^2\right|\mathcal{F}_k\right] \frac{\left\|z_{N,k}^\prime\right\|^\beta \cdot\mathbb{E}\left[\left.\left\|\varepsilon_{k+1}\right\|^\beta\right|\mathcal{F}_k\right]}{\varepsilon^{\beta-2} \cdot \mathbb{E}\left[\left.\left(z_{N,k}^\prime \varepsilon_{k+1}\right)^2\right|\mathcal{F}_k\right]}.\label{eq:zpsqe0}
\end{align}
and
\begin{align}\label{eq:zpsq}
    &\mathbb{E}\left[\left.\left(z_{N,k}^\prime \varepsilon_{k+1}\right)^2\right|\mathcal{F}_k\right]\nonumber\\
    &=\mathbb{E}\Big[\big(z_{N,k}^\prime(1) \varepsilon_{k+1}(1)+z_{N,k}^\prime(2) \varepsilon_{k+1}(2)+\cdots\nonumber\\
    &~~~~~~~~+z_{N,k}^\prime(n) \varepsilon_{k+1}(n)\big)^2\Big|\mathcal{F}_k\Big]\nonumber\\
    &=\mathbb{E}\Big[\sum_{i=1}^n \left(z_{N,k}^\prime(i) \varepsilon_{k+1}(i)\right)^2\nonumber\\
    &~~~+2\sum_{1\leq i<j\leq n} z_{N,k}^\prime(i) \varepsilon_{k+1}(i) z_{N,k}^\prime(j) \varepsilon_{k+1}(j)\Big|\mathcal{F}_k\Big]\nonumber\\
    &=\sigma^2\left\|z_{N,k}^\prime\right\|^2.
\end{align}

By substituting (\ref{eq:zpsq}) into (\ref{eq:zpsqe0}), we have that
\begin{align}\label{eq:zpesqe}
    &\sum_{k=1}^N \mathbb{E}\left[\left.\left(z_{N,k}^\prime \varepsilon_{k+1}\right)^2 \mathbb{I}\left\{\left|z_{N,k}^\prime \varepsilon_{k+1}\right|>\varepsilon\right\}\right|\mathcal{F}_k\right] \nonumber\\
    &\leq \sum_{k=1}^N \mathbb{E}\left[\left.\left(z_{N,k}^\prime \varepsilon_{k+1}\right)^2\right|\mathcal{F}_k\right] \frac{\left\|z_{N,k}^\prime\right\|^\beta \cdot\mathbb{E}\left[\left.\left\|\varepsilon_{k+1}\right\|^\beta\right|\mathcal{F}_k\right]}{\varepsilon^{\beta-2} \cdot \sigma^2\left\|z_{N,k}^\prime\right\|^2} \nonumber\\
    &\leq \Big(\max_{1\leq k \leq N} \left\|z_{N,k}^\prime\right\|^{\beta-2}\Big) \cdot \sum_{k=1}^N \mathbb{E}\left[\left.\left(z_{N,k}^\prime \varepsilon_{k+1}\right)^2\right|\mathcal{F}_k\right] \nonumber\\
    &~~~\cdot \frac{ \sup_{k\geq 1} \mathbb{E}\left[\left.\left\|\varepsilon_{k+1}\right\|^{\beta}\right|\mathcal{F}_k\right]}{\varepsilon^{\beta-2} \cdot \sigma^2} \nonumber\\
    &= O\left(\max_{1\leq k \leq N} \left\|z_{N,k}^\prime\right\|^{\beta-2} \cdot \sum_{k=1}^N \mathbb{E}\left[\left.\left(z_{N,k}^\prime \varepsilon_{k+1}\right)^2\right|\mathcal{F}_k\right]\right)\nonumber\\
    &~~~\mathop{\longrightarrow} \limits_{N \rightarrow \infty}^{\mathbb{P}} 0
\end{align}
where for last limit taking place Assumption \ref{assum:1}, (\ref{eq:zpto0}), and (\ref{eq:zpesq}) are applied.

Combining (\ref{eq:condition1}), (\ref{eq:zpesq}), and (\ref{eq:zpesqe}), by Lemma \ref{lem:clt} it yields that
\begin{align}\label{eq:zpasy}
    \sum_{k=1}^N z_{N,k}^\prime \varepsilon_{k+1}
    \mathop{\longrightarrow} \limits_{N \rightarrow \infty}^\mathcal{D}
    \mathcal{N}\left(0, \sigma^2 \|v\|^2\right).
\end{align}

Denote the set $A_N\triangleq \left\{\omega :z_{N,k}(\omega)\neq z_{N,k}^\prime(\omega), \text{ for some } 1\leq k \leq N\right\}$. From (\ref{eq:zzp}) it yields that $\mathbb{P}\left\{A_N\right\}\mathop{\longrightarrow} \limits_{N \rightarrow \infty} 0$. For any $\delta>0$, noting that $$\left\{\left|\sum_{k=1}^N z_{N,k} \varepsilon_{k+1}-\sum_{k=1}^N z_{N,k}^\prime \varepsilon_{k+1}\right|>\delta\right\}\subseteq A_N,$$
we have
\begin{align}\label{eq:pzzp}
    &\lim_{N\rightarrow \infty} \mathbb{P}\left\{\left|\sum_{k=1}^N z_{N,k} \varepsilon_{k+1}-\sum_{k=1}^N z_{N,k}^\prime \varepsilon_{k+1}\right|>\delta\right\}\nonumber\\
    &\leq \lim_{N\rightarrow \infty} \mathbb{P}\left\{A_N\right\}=0.
\end{align}

From (\ref{eq:zpasy}) and (\ref{eq:pzzp}), by Slutsky's theorem it follows that (\ref{eq:zasy}) holds, which together with (\ref{eq:zasy3})--(\ref{eq:zasy1}) yields the conclusion of Lemma \ref{lem:LSasy}. This finishes the proof.

\subsection{Proof of Theorem \ref{th:asydisx}}\label{app:thasydisx}

\textcolor{black}{We first consider the case ${\rm rank}(\Theta)=r<n$}. Note that
\begin{align}\label{eq:trans}
    &\left(I_n \otimes C_{N} U_{N+1}^{(1)} {U_{N+1}^{(1)}}^\top\right) {\rm vec}\left(X_{N+1}-\Theta\right)\nonumber\\
    &={\rm vec}\left(C_{N} U_{N+1}^{(1)} {U_{N+1}^{(1)}}^\top \left(X_{N+1}-\Theta\right)\right).
\end{align}

In the following, we first analyze $C_{N} U_{N+1}^{(1)} {U_{N+1}^{(1)}}^\top (X_{N+1}-\Theta)$.

Recall that the SVD of the RLS estimate $\Theta_{N+1}$ has the following form,
\begin{align}\label{eq:xn}
\Theta_{N+1}=U_{N+1} \left[
\begin{matrix}
\Sigma_{N+1}^{(1)} & 0\\
0 & \Sigma_{N+1}^{(2)}
\end{matrix}
\right] V_{N+1}^\top
\end{align}
where
\begin{align}
    \Sigma_{N+1}^{(1)}&\triangleq {\rm diag} \left\{\sigma_1\left(\Theta_{N+1}\right),\cdots, \sigma_{r}\left(\Theta_{N+1}\right)\right\},\nonumber\\
    \Sigma_{N+1}^{(2)}&\triangleq \left[
\begin{smallmatrix}
    {\rm diag} \left\{\sigma_{r+1}\left(\Theta_{N+1}\right),\cdots, \sigma_{n}\left(\Theta_{N+1}\right)\right\}\\
    0_{(d-n)\times (n-r)}
\end{smallmatrix}
\right]\in \mathbb{R}^{(d-r)\times (n-r)}.\nonumber
\end{align}

Set
\begin{align}\label{eq:dxn}
    \Delta X_{N+1}
    \triangleq \Theta_{N+1}-X_{N+1},
\end{align}
then
$X_{N+1}=\Theta_{N+1}-\Delta X_{N+1}.$

The left and right singular value matrices $U_{N+1}\in \mathbb{R}^{d \times d}$ and $V_{N+1} \in \mathbb{R}^{n \times n}$ can be formulated as $U_{N+1}=\left[U_{N+1}^{(1)}~U_{N+1}^{(2)}\right]$ with $U_{N+1}^{(1)}\in \mathbb{R}^{d \times r},~U_{N+1}^{(2)}\in \mathbb{R}^{d \times (d-r)}$ and $V_{N+1}=\left[V_{N+1}^{(1)}~V_{N+1}^{(2)}\right]$ with $V_{N+1}^{(1)}\in \mathbb{R}^{n \times r},~V_{N+1}^{(2)}\in \mathbb{R}^{n \times (n-r)}$, respectively.

By the orthogonality of $U_{N+1}$ and $V_{N+1}$, one has
\begin{align}\label{eq:thetax}
    &{U_{N+1}^{(1)}}^\top \Theta_{N+1}\nonumber\\
    &={U_{N+1}^{(1)}}^\top \left[U_{N+1}^{(1)}~U_{N+1}^{(2)}\right] \left[
    \begin{matrix}
        \Sigma_{N+1}^{(1)} & 0\\
        0 & \Sigma_{N+1}^{(2)}
    \end{matrix}
    \right] V_{N+1}^\top\nonumber\\
    &=\left[
    \begin{matrix}
        I_r & 0
    \end{matrix}
    \right] \left[
    \begin{matrix}
        \Sigma_{N+1}^{(1)} & 0\\
        0 & \Sigma_{N+1}^{(2)}
    \end{matrix}
    \right]
    \left[
    \begin{matrix}
        {V_{N+1}^{(1)}}^\top\\
        {V_{N+1}^{(2)}}^\top
    \end{matrix}
    \right]\nonumber\\
    &=\Sigma_{N+1}^{(1)} {V_{N+1}^{(1)}}^\top
    ={U_{N+1}^{(1)}}^\top \left(X_{N+1}+\Delta X_{N+1}\right).
\end{align}

Note that the RLS estimates can also be written as 
\begin{align}
    \Theta_{N+1}=P_{N+1}\left(\sum_{k=1}^N \varphi_k y_{k+1}^\top + P_1^{-1} \Theta_1 \right),
\end{align}
which combining with (\ref{eq:thetax}) yields that
\begin{align}
    &{U_{N+1}^{(1)}}^\top \left(X_{N+1}+\Delta X_{N+1}-\Theta\right)
    ={U_{N+1}^{(1)}}^\top \left(\Theta_{N+1}-\Theta\right)\nonumber\\
    &={U_{N+1}^{(1)}}^\top \left(P_{N+1}\sum_{k=1}^N \varphi_k y_{k+1}^\top+P_{N+1} P_1^{-1} \Theta_1-\Theta\right).\label{eq97}
\end{align}

By the definition of $P_{N+1}$ and (\ref{eq97}) and noting $y_{k+1}=\Theta^\top \varphi_k + \varepsilon_{k+1}$, we have
\begin{align}
    P_{N+1}\sum_{k=1}^N \varphi_k \varphi_k^\top
    &=P_{N+1}\left(\sum_{k=1}^N \varphi_k \varphi_k^\top+P_1^{-1}\right)-P_{N+1} P_1^{-1}\nonumber\\
    &=I_d-P_{N+1} P_1^{-1}
\end{align}
and 
\begin{align}
    &{U_{N+1}^{(1)}}^\top \left(X_{N+1}+\Delta X_{N+1}-\Theta\right) \nonumber\\
    &={U_{N+1}^{(1)}}^\top \Big(\Theta-P_{N+1}  P_1^{-1} \Theta + P_{N+1} \sum_{k=1}^N \varphi_k \varepsilon_{k+1}^\top\nonumber\\&~~~+P_{N+1} P_1^{-1} \Theta_1-\Theta\Big)\nonumber\\
    &={U_{N+1}^{(1)}}^\top P_{N+1} \sum_{k=1}^N \varphi_k \varepsilon_{k+1}^\top +{U_{N+1}^{(1)}}^\top P_{N+1} P_1^{-1}\left(\Theta_1-\Theta\right).
\end{align}

Multiplying both sides of the above equation by $C_{N} U_{N+1}^{(1)}$ from the left, we obtain that
\begin{align}\label{eq:cu}
    &C_{N} U_{N+1}^{(1)} {U_{N+1}^{(1)}}^\top \left(X_{N+1}+\Delta X_{N+1}-\Theta\right)\nonumber\\
    &=C_{N} U_{N+1}^{(1)} {U_{N+1}^{(1)}}^\top P_{N+1} \sum_{k=1}^N \varphi_k \varepsilon_{k+1}^\top\nonumber\\
    &~~~+C_{N} U_{N+1}^{(1)} {U_{N+1}^{(1)}}^\top P_{N+1}P_1^{-1}\left(\Theta_1-\Theta\right).
\end{align}

By (\ref{eq:trans}) and (\ref{eq:cu}), it follows that
\begin{align}\label{eq:trans00}
    &\left(I_n \otimes C_{N} U_{N+1}^{(1)} {U_{N+1}^{(1)}}^\top\right) {\rm vec}\left(X_{N+1}-\Theta\right) \nonumber\\
    &={\rm vec}\left(Q_N \cdot R_N \cdot \left(\sum_{k=1}^N \varphi_k \varphi_k^\top\right)^{-1/2} \sum_{k=1}^N \varphi_k \varepsilon_{k+1}^\top\right)\nonumber\\
    &~~~+{\rm vec}\left(T_N\right)-{\rm vec}\left(Z_N\right)\nonumber\\
    &=\left(I_n \otimes \left(Q_N \cdot R_N\right)\right){\rm vec}\left(\left(\sum_{k=1}^N \varphi_k \varphi_k^\top\right)^{-1/2} \sum_{k=1}^N \varphi_k \varepsilon_{k+1}^\top\right)\nonumber\\
    &~~~+{\rm vec}\left(T_N\right)-{\rm vec}\left(Z_N\right)
\end{align}
where
\begin{align*}
    Q_{N} &\triangleq C_{N} U_{N+1}^{(1)} {U_{N+1}^{(1)}}^\top \left(\sum_{k=1}^N \varphi_k \varphi_k^\top\right)^{-1/2},\\
    R_N &\triangleq \left(\sum_{k=1}^N \varphi_k \varphi_k^\top\right)^{1/2} P_{N+1} \left(\sum_{k=1}^N \varphi_k \varphi_k^\top\right)^{1/2},\\
    T_N &\triangleq C_{N} U_{N+1}^{(1)} {U_{N+1}^{(1)}}^\top P_{N+1}P_1^{-1}\left(\Theta_1-\Theta\right),\\
    Z_N & \triangleq C_{N} U_{N+1}^{(1)} {U_{N+1}^{(1)}}^\top \Delta X_{N+1}.
\end{align*}


For $Q_N$ in (\ref{eq:trans00}), we have
\begin{align}\label{eq:QN0}
    Q_N
    &=C_{N} \left(U_{N+1}^{(1)} {U_{N+1}^{(1)}}^\top - U^{(1)} {U^{(1)}}^\top\right) \left(\sum_{k=1}^N \varphi_k \varphi_k^\top\right)^{-1/2} \nonumber\\
    &~~~+ C_{N} U^{(1)} {U^{(1)}}^\top \left(\sum_{k=1}^N \varphi_k \varphi_k^\top\right)^{-1/2}.
\end{align}
By Assumption \ref{assum:6}, the second and the first terms on the right-hand side of (\ref{eq:QN0}) satisfy
\begin{align}\label{eq:QN2}
    C_{N} U^{(1)} {U^{(1)}}^\top \left(\sum_{k=1}^N \varphi_k \varphi_k^\top\right)^{-1/2} \mathop{\longrightarrow} \limits_{N \rightarrow \infty}^{\mathbb{P}} M
\end{align}
and
\begin{align}\label{eq:QN1-1}
    &\left\|C_{N} \left(U_{N+1}^{(1)} {U_{N+1}^{(1)}}^\top - U^{(1)} {U^{(1)}}^\top\right) \left(\sum_{k=1}^N \varphi_k \varphi_k^\top\right)^{-1/2}\right\|\nonumber\\
    &\leq \left\|C_{N}\right\| \left\|U_{N+1}^{(1)} {U_{N+1}^{(1)}}^\top - U^{(1)} {U^{(1)}}^\top\right\| \left\|\left(\sum_{k=1}^N \varphi_k \varphi_k^\top\right)^{-1/2}\right\|\nonumber\\
    &=O_p\left(\left\|C_{N}\right\| \sqrt{\frac{\log\lambda_{\max}(N)}{\lambda_{\min}(N)}} \lambda_{\min}^{-1/2}\left\{\sum_{k=1}^N \varphi_k \varphi_k^\top\right\}\right)\nonumber\\
    &=O_p\left(\left\|C_{N}\right\| \frac{\sqrt{\log\lambda_{\max}(N)}}{\lambda_{\min}(N)}\right)\mathop{\longrightarrow} \limits_{N \rightarrow \infty}^{\mathbb{P}}0,
\end{align}
respectively.

Hence it follows that
\begin{align}\label{eq:QN1}
    C_{N} \left(U_{N+1}^{(1)} {U_{N+1}^{(1)}}^\top - U^{(1)} {U^{(1)}}^\top\right) \left(\sum_{k=1}^N \varphi_k \varphi_k^\top\right)^{-1/2} \mathop{\longrightarrow} \limits_{N \rightarrow \infty}^{\mathbb{P}} 0
\end{align}
and
\begin{align}\label{eq:QN}
    Q_{N} = C_{N} U_{N+1}^{(1)} {U_{N+1}^{(1)}}^\top \left(\sum_{k=1}^N \varphi_k \varphi_k^\top\right)^{-1/2} \mathop{\longrightarrow} \limits_{N \rightarrow \infty}^{\mathbb{P}} M.
\end{align}

For $R_N$ in (\ref{eq:trans00}), by the definition of $P_{N+1}$ it can be directly verified that
\begin{align}\label{eq:cu2}
R_N = \left(\sum_{k=1}^N \varphi_k \varphi_k^\top\right)^{1/2} P_{N+1} \left(\sum_{k=1}^N \varphi_k \varphi_k^\top\right)^{1/2} \mathop{\longrightarrow} \limits_{N \rightarrow \infty}^{\mathbb{P}} I_d.
\end{align}

For $T_N$ in (\ref{eq:trans00}), direct calculation leads to
\begin{align}
T_N
    &=Q_N \cdot R_N \cdot \left(\sum_{k=1}^N \varphi_k \varphi_k^\top\right)^{-1/2} P_1^{-1}\left(\Theta_1-\Theta\right),
\end{align}
from which and noting $\lambda_{\min}(N)\mathop{\longrightarrow} \limits_{N \rightarrow \infty}^\mathbb{P} \infty$,
\begin{align}
    \left\|T_N\right\|
    &\leq \left\|Q_N\right\| \cdot \left\|R_N\right\| \cdot \left\|\left(\sum_{k=1}^N \varphi_k \varphi_k^\top\right)^{-1/2}\right\| \nonumber\\
    &~~~\cdot \left\|P_1^{-1}\left(\Theta_1-\Theta\right)\right\|\nonumber\\
    &=O_{p}\left(\lambda_{\min}^{-1/2}(N)\right)\mathop{\longrightarrow} \limits_{N \rightarrow \infty}^\mathbb{P} 0.\label{eq109}
\end{align}


Next, we analyze $Z_N$ in (\ref{eq:trans00}). By Theorem \ref{th:rank}, there exists an $\omega$-set $\Omega_0$ with $\mathbb{P}\{\Omega_0\}=1$ such that for any fixed $\omega\in\Omega_0$, $\mathrm{rank}(X_{N+1}(\omega))=\widehat{r}_{N}(\omega)=r$ for all $N$ large enough. In the following analysis, we focus on a fixed $\omega\in\Omega_0$. By the definition of $\Delta X_{N+1}$,
\begin{align}\label{eq:dxn1}
    &\Delta X_{N+1}
    =U_{N+1} \nonumber\\
    &\cdot\left[
    \begin{smallmatrix}
        {\rm diag}\left\{\sigma_i(\Theta_{N+1}), i=1,\cdots,\widehat{r}_N\right\} & 0_{\widehat{r}_N \times (n-\widehat{r}_N)}\\
        0_{(n-\widehat{r}_N) \times \widehat{r}_N} & {\rm diag}\left\{\sigma_{i}(\Theta_{N+1}), i=\widehat{r}_N+1,\cdots,n\right\}\\
        0_{(d-n)\times \widehat{r}_N} & 0_{(d-n)\times (n-\widehat{r}_N})
    \end{smallmatrix}
    \right] V_{N+1}^\top\nonumber\\
    &-
    U_{N+1} \nonumber\\
    &\cdot \left[
    \begin{smallmatrix}
        {\rm diag}\left\{\sigma_i(\Theta_{N+1})-\lambda_N w_{N+1}(i), i=1,\cdots,\widehat{r}_N\right\} & 0_{\widehat{r}_N \times (n-\widehat{r}_N)}\\
        0_{(d-\widehat{r}_N)\times \widehat{r}_N} & 0_{(d-\widehat{r}_N)\times (n-\widehat{r}_N})
    \end{smallmatrix}
    \right] V_{N+1}^\top\nonumber\\
    &=U_{N+1} \nonumber\\
    &\cdot \left[
    \begin{smallmatrix}
        {\rm diag}\left\{\lambda_N w_{N+1}(i), i=1,\cdots,r\right\} & 0_{r \times (n-r)}\\
        0_{(n-r)\times r} & {\rm diag}\left\{\sigma_{i}(\Theta_{N+1}), i=r,\cdots,n\right\}\\
        0_{(d-n)\times r} & 0_{(d-n)\times (n-r)}
    \end{smallmatrix}
    \right] V_{N+1}^\top.
\end{align}
Noting ${U_{N+1}^{(1)}}^\top U_{N+1}=\left[\begin{matrix} I_r & 0_{r\times (d-r)} \end{matrix}\right]$ and (\ref{eq:dxn1}), we obtain that
\begin{align}\label{eq:zn0}
    &{U_{N+1}^{(1)}}^\top \Delta X_{N+1}\nonumber\\
    &=\left[
    \begin{smallmatrix}
    {\rm diag}\left\{\lambda_N w_{N+1}(1), \cdots,\lambda_N w_{N+1}(r)\right\} & \textcolor{black}{0_{r\times (n-r)}
    }
    \end{smallmatrix}
    \right] V_{N+1}^\top
\end{align}
and by noting $\mathrm{rank}(\Theta)=r$ and the definitions of $w_{N+1}(i),~i=1,\cdots,r$,
\begin{align}\label{eq:udxn1}
    \left\|{U_{N+1}^{(1)}}^\top\Delta X_{N+1}\right\|
    &=O\left(\lambda_N\right).
\end{align}
Since (\ref{eq:udxn1}) is obtained for any fixed $\omega\in\Omega_0$, we can further obtain that
\begin{align}\label{eq:udxn2}
    \left\|{U_{N+1}^{(1)}}^\top\Delta X_{N+1}\right\|
    &=O\left(\lambda_N\right),~~\mathrm{a.s.}\nonumber\\
    &=O_p\left(\lambda_N\right).
\end{align}

From (\ref{eq:udxn2}) and (\ref{eq:cnlambdan}) in Assumption \ref{assum:6}, we know that
\begin{align}\label{eq:zn}
    \left\|Z_N\right\|
    &= \left\|C_{N} U_{N+1}^{(1)} {U_{N+1}^{(1)}}^\top \Delta X_{N+1}\right\|\nonumber\\
    &\leq \left\|C_{N}\right\| \left\|U_{N+1}^{(1)}\right\| \left\|{U_{N+1}^{(1)}}^\top\Delta X_{N+1}\right\|\nonumber\\
    &= O_p\left(\lambda_N \left\|C_{N}\right\|\right)
    \mathop{\longrightarrow} \limits_{N \rightarrow \infty}^\mathbb{P} 0.
\end{align}


Combining (\ref{eq:trans00}), (\ref{eq:QN}), (\ref{eq:cu2}), (\ref{eq109}), and (\ref{eq:zn}), and noting Lemma \ref{lem:LSasy}, we have that
\begin{align}
    \left(I_n \otimes C_{N} U_{N+1}^{(1)} {U_{N+1}^{(1)}}^\top\right) {\rm vec}\left(X_{N+1}-\Theta\right)\nonumber\\
    \mathop{\longrightarrow} \limits_{N \rightarrow \infty}^\mathcal{D} \mathcal{N}\left(0_{dn}, \sigma^2 I_n \otimes \left(M M^\top \right)\right).\nonumber
\end{align}

Next, we consider the case ${\rm rank}(\Theta)=n$. Note that
\begin{align}\label{eq:asyfull}
    \left(I_n \otimes C_{N}\right) {\rm vec}\left(X_{N+1}-\Theta\right)
    ={\rm vec}\left(C_{N}  \left(X_{N+1}-\Theta\right)\right).
\end{align}

In the following we analyze $C_{N} \left(X_{N+1}-\Theta\right)$. Since $X_{N+1}=\Theta_{N+1}-\Delta X_{N+1}$, we have
\begin{align}\label{eq:asyfull1}
    &C_{N} \left(X_{N+1}-\Theta\right)\nonumber\\
    &=C_{N}  \left(\Theta_{N+1}-\Theta\right)-C_{N}  \Delta X_{N+1}\nonumber\\
    &=C_{N} \left(\sum_{k=1}^N \varphi_k \varphi_k^\top\right)^{-1/2} \cdot \left(\sum_{k=1}^N \varphi_k \varphi_k^\top\right)^{1/2}\left(\Theta_{N+1}-\Theta\right)\nonumber\\
    &~~~-C_{N} \Delta X_{N+1}.
\end{align}
By Theorem \ref{th:rank}, there exists an $\omega$-set $\Omega_0$ with $\mathbb{P}\{\Omega_0\}=1$ such that for any fixed $\omega\in\Omega_0$, $\mathrm{rank}(X_{N+1}(\omega))=\widehat{r}_{N}(\omega)=n$ for all $N$ large enough. We focus on a fixed $\omega\in\Omega_0$. Similar to (\ref{eq:dxn1}), we obtain
\begin{align}
    &\Delta X_{N+1}\nonumber\\
    &=U_{N+1} \left[
    \begin{smallmatrix}
        {\rm diag}\left\{\lambda_N w_{N+1}(1), \cdots,\lambda_N w_{N+1}(n)\right\}; 0_{(d-n)\times n}
    \end{smallmatrix}
    \right]
    V_{N+1}^\top,\nonumber
\end{align}
and by noting $\mathrm{rank}(\Theta)=n$ and the definitions of $w_{N+1}(i),~i=1,\cdots,n$,
\begin{align}\label{eq:dxnorm}
    \left\|\Delta X_{N+1}\right\|
    &=O\left(\lambda_N\right),~~\mathrm{a.s.}\nonumber\nonumber\\
    &=O_p\left(\lambda_N\right).
\end{align}

By (\ref{eq:dxnorm}) and (\ref{eq:cnlambdanfull}) in Assumption \ref{assum:6}, we have
\begin{align}\label{eq:cndxnorm}
    \left\|C_{N} \Delta X_{N+1}\right\|
    &\leq \left\|C_{N}\right\| \left\|\Delta X_{N+1}\right\|\nonumber\\
    &=O_p\left(\lambda_N \left\|C_{N}\right\|\right)
    \mathop{\longrightarrow} \limits_{N \rightarrow \infty}^\mathbb{P} 0,
\end{align}
and
\begin{align}\label{eq:cnfull}
    C_{N} \left(\sum_{k=1}^N \varphi_k \varphi_k^\top\right)^{-1/2} \mathop{\longrightarrow} \limits^{\mathbb{P}}_{N \rightarrow \infty} I_d.
\end{align}

Combining (\ref{eq:asyfull}), (\ref{eq:asyfull1}),  (\ref{eq:cndxnorm}), and (\ref{eq:cnfull}), and following the same procedure as for the case $\mathrm{rank}(\Theta)<n$, we can prove that
\begin{align}
\left(I_n \otimes C_{N}\right) {\rm vec}\left(X_{N+1}-\Theta\right)
\mathop{\longrightarrow} \limits_{N \rightarrow \infty}^\mathcal{D} \mathcal{N}(0_{dn}, \sigma^2 I_{dn}).
\end{align}

This finishes the proof.



\bibliographystyle{plain}  
\bibliography{Reference}

\begin{thebibliography}{10}

\bibitem{alvarez2017compression}
Jose~M Alvarez and Mathieu Salzmann.
\newblock Compression-aware training of deep networks.
\newblock In {\em Conference on Neural Information Processing Systems}, 2017.

\bibitem{aastrom1973self}
Karl~Johan {\AA}str{\"o}m and Bj{\"o}rn Wittenmark.
\newblock On self tuning regulators.
\newblock {\em Automatica}, 9:185--199, 1973.

\bibitem{breiman1996heuristics}
Leo Breiman.
\newblock Heuristics of instability and stabilization in model selection.
\newblock {\em The Annals of Statistics}, 24:2350--2383, 1996.

\bibitem{buchanan2005damped}
Aeron~M Buchanan and Andrew~W Fitzgibbon.
\newblock Damped newton algorithms for matrix factorization with missing data.
\newblock In {\em Conference on Computer Vision and Pattern Recognition}, 2005.

\bibitem{bunea2011optimal}
Florentina Bunea, Yiyuan She, and Marten~H. Wegkamp.
\newblock {Optimal selection of reduced rank estimators of high-dimensional
  matrices}.
\newblock {\em The Annals of Statistics}, 39:1282 -- 1309, 2011.

\bibitem{cai2010singular}
Jian-Feng Cai, Emmanuel~J Cand{\`e}s, and Zuowei Shen.
\newblock A singular value thresholding algorithm for matrix completion.
\newblock {\em SIAM Journal on Optimization}, 20:1956--1982, 2010.

\bibitem{candes2011tight}
Emmanuel~J Cand{\`e}s and Yaniv Plan.
\newblock Tight oracle inequalities for low-rank matrix recovery from a minimal
  number of noisy random measurements.
\newblock {\em IEEE Transactions on Information Theory}, 57:2342--2359, 2011.

\bibitem{candes2012exact}
Emmanuel~J Cand{\`e}s and Benjamin Recht.
\newblock Exact matrix completion via convex optimization.
\newblock {\em Communications of the ACM}, 55:111--119, 2012.

\bibitem{cao2023identification}
Wenqi Cao, Giorgio Picci, and Anders Lindquist.
\newblock Identification of low rank vector processes.
\newblock {\em Automatica}, 151:110938, 2023.

\bibitem{chen2012identification}
Han-Fu Chen and Lei Guo.
\newblock {\em Identification and stochastic adaptive control}.
\newblock Birkhäuser Boston, MA, 1991.

\bibitem{chen2013reduced}
Kun Chen, Hongbo Dong, and Kung-Sik Chan.
\newblock Reduced rank regression via adaptive nuclear norm penalization.
\newblock {\em Biometrika}, 100:901--920, 2013.

\bibitem{chen2012sparse}
Lisha Chen and Jianhua~Z Huang.
\newblock Sparse reduced-rank regression for simultaneous dimension reduction
  and variable selection.
\newblock {\em Journal of the American Statistical Association},
  107:1533--1545, 2012.

\bibitem{chen2021drone}
Patrick Chen, Hsiang-Fu Yu, Inderjit Dhillon, and Cho-Jui Hsieh.
\newblock Drone: Data-aware low-rank compression for large \text{NLP} models.
\newblock In {\em Conference on Neural Information Processing Systems}, 2021.

\bibitem{chen2023joint}
Shaowu Chen, Jiahao Zhou, Weize Sun, and Lei Huang.
\newblock Joint matrix decomposition for deep convolutional neural networks
  compression.
\newblock {\em Neurocomputing}, 516:11--26, 2023.

\bibitem{chizhik2002keyholes}
Dmitry Chizhik, Gerard~J Foschini, Michael~J Gans, and Reinaldo~A Valenzuela.
\newblock Keyholes, correlations, and capacities of multielement transmit and
  receive antennas.
\newblock {\em IEEE Transactions on Wireless Communications}, 1:361--368, 2002.

\bibitem{chow2012probability}
Yuan~Shih Chow and Henry Teicher.
\newblock {\em Probability theory: independence, interchangeability,
  martingales}.
\newblock Springer Science \& Business Media, 2012.

\bibitem{cui2022channel}
Mingyao Cui and Linglong Dai.
\newblock Channel estimation for extremely large-scale \text{MIMO}: Far-field
  or near-field?
\newblock {\em IEEE Transactions on Communications}, 70:2663--2677, 2022.

\bibitem{davenport2016overview}
Mark~A Davenport and Justin Romberg.
\newblock An overview of low-rank matrix recovery from incomplete observations.
\newblock {\em IEEE Journal of Selected Topics in Signal Processing},
  10:608--622, 2016.

\bibitem{denil2013predicting}
Misha Denil, Babak Shakibi, Laurent Dinh, Marc'Aurelio Ranzato, and Nando
  De~Freitas.
\newblock Predicting parameters in deep learning.
\newblock In {\em Conference on Neural Information Processing Systems}, 2013.

\bibitem{dogariu2020efficient}
Laura-Maria Dogariu, Constantin Paleologu, Jacob Benesty, and Silviu
  Ciochin{\u{a}}.
\newblock An efficient \text{Kalman} filter for the identification of low-rank
  systems.
\newblock {\em Signal Processing}, 166:107239, 2020.

\bibitem{elisei2019recursive}
Camelia Elisei-Iliescu, Constantin Paleologu, Jacob Benesty, Cristian Stanciu,
  Cristian Anghel, and Silviu Ciochin{\u{a}}.
\newblock Recursive least-squares algorithms for the identification of low-rank
  systems.
\newblock {\em IEEE/ACM Transactions on Audio, Speech, and Language
  Processing}, 27:903--918, 2019.

\bibitem{eriksson2010efficient}
Anders Eriksson and Anton Van Den~Hengel.
\newblock Efficient computation of robust low-rank matrix approximations in the
  presence of missing data using the \text{$L_1$} norm.
\newblock In {\em Conference on Computer Vision and Pattern Recognition}, 2010.

\bibitem{fazel2004rank}
Maryam Fazel, Haitham Hindi, and Stephen Boyd.
\newblock Rank minimization and applications in system theory.
\newblock In {\em American Control Conference}, 2004.

\bibitem{fazel2001rank}
Maryam Fazel, Haitham Hindi, and Stephen~P Boyd.
\newblock A rank minimization heuristic with application to minimum order
  system approximation.
\newblock In {\em American Control Conference}, 2001.

\bibitem{fazel2013hankel}
Maryam Fazel, Ting~Kei Pong, Defeng Sun, and Paul Tseng.
\newblock Hankel matrix rank minimization with applications to system
  identification and realization.
\newblock {\em SIAM Journal on Matrix Analysis and Applications}, 34:946--977,
  2013.

\bibitem{golub2013matrix}
Gene~H Golub and Charles~F Van~Loan.
\newblock {\em Matrix computations}.
\newblock Johns Hopkins University Press, 2013.

\bibitem{gu2014weighted}
Shuhang Gu, Lei Zhang, Wangmeng Zuo, and Xiangchu Feng.
\newblock Weighted nuclear norm minimization with application to image
  denoising.
\newblock In {\em Conference on Computer Vision and Pattern Recognition}, 2014.

\bibitem{halko2011finding}
Nathan Halko, Per-Gunnar Martinsson, and Joel~A Tropp.
\newblock Finding structure with randomness: Probabilistic algorithms for
  constructing approximate matrix decompositions.
\newblock {\em SIAM Review}, 53:217--288, 2011.

\bibitem{hansen2022probability}
B.~Hansen.
\newblock {\em Probability and Statistics for Economists}.
\newblock Princeton University Press, 2022.

\bibitem{matrixanalysis2012}
Roger~A Horn and Charles~R Johnson.
\newblock {\em Matrix analysis}.
\newblock Cambridge University Press, 2012.

\bibitem{hsu2022language}
Yen-Chang Hsu, Ting Hua, Sungen Chang, Qian Lou, Yilin Shen, and Hongxia Jin.
\newblock Language model compression with weighted low-rank factorization.
\newblock In {\em International Conference on Learning Representations}, 2022.

\bibitem{hu2012fast}
Yao Hu, Debing Zhang, Jieping Ye, Xuelong Li, and Xiaofei He.
\newblock Fast and accurate matrix completion via truncated nuclear norm
  regularization.
\newblock {\em IEEE Transactions on Pattern Analysis and Machine Intelligence},
  35:2117--2130, 2012.

\bibitem{Reweighted2019}
Yan Huang, Guisheng Liao, Yijian Xiang, Lei Zhang, Jie Li, and Arye Nehorai.
\newblock Low-rank approximation via generalized reweighted iterative nuclear
  and \text{Frobenius} norms.
\newblock {\em IEEE Transactions on Image Processing}, 29:2244--2257, 2019.

\bibitem{izenman1975reduced}
Alan~Julian Izenman.
\newblock Reduced-rank regression for the multivariate linear model.
\newblock {\em Journal of Multivariate Analysis}, 5:248--264, 1975.

\bibitem{jaderberg2014speeding}
Max Jaderberg, Andrea Vedaldi, and Andrew Zisserman.
\newblock Speeding up convolutional neural networks with low rank expansions.
\newblock In {\em Conference on Neural Information Processing Systems}, 2019.

\bibitem{ke2005robust}
Qifa Ke and Takeo Kanade.
\newblock Robust \text{$L_1$} norm factorization in the presence of outliers
  and missing data by alternative convex programming.
\newblock In {\em Conference on Computer Vision and Pattern Recognition}, 2005.

\bibitem{lai1981consistency}
Tze~Leung Lai and Herbert Robbins.
\newblock Consistency and asymptotic efficiency of slope estimates in
  stochastic approximation schemes.
\newblock {\em Zeitschrift f{\"u}r Wahrscheinlichkeitstheorie und Verwandte
  Gebiete}, 56:329--360, 1981.

\bibitem{lai1982least}
Tze~Leung Lai and Ching~Zong Wei.
\newblock Least squares estimates in stochastic regression models with
  applications to identification and control of dynamic systems.
\newblock {\em The Annals of Statistics}, 10:154--166, 1982.

\bibitem{li2022high}
Junlin Li.
\newblock High-dimensional dynamic systems identification with additional
  constraints.
\newblock {\em Communications in Statistics-Theory and Methods}, 51:5204--5225,
  2022.

\bibitem{li2023losparse}
Yixiao Li, Yifan Yu, Qingru Zhang, Chen Liang, Pengcheng He, Weizhu Chen, and
  Tuo Zhao.
\newblock Losparse: Structured compression of large language models based on
  low-rank and sparse approximation.
\newblock In {\em International Conference on Machine Learning}, 2023.

\bibitem{lu2014generalized}
Canyi Lu, Jinhui Tang, Shuicheng Yan, and Zhouchen Lin.
\newblock Generalized nonconvex nonsmooth low-rank minimization.
\newblock In {\em Conference on Computer Vision and Pattern Recognition}, pages
  4130--4137, 2014.

\bibitem{negahban2011estimation}
Sahand Negahban and Martin~J. Wainwright.
\newblock {Estimation of (near) low-rank matrices with noise and
  high-dimensional scaling}.
\newblock {\em The Annals of Statistics}, 39:1069 -- 1097, 2011.

\bibitem{negahban2009unified}
Sahand Negahban, Bin Yu, Martin~J Wainwright, and Pradeep Ravikumar.
\newblock A unified framework for high-dimensional analysis of $m$-estimators
  with decomposable regularizers.
\newblock In {\em Conference on Neural Information Processing Systems}, 2009.

\bibitem{paleologu2018linear}
Constantin Paleologu, Jacob Benesty, and Silviu Ciochin{\u{a}}.
\newblock Linear system identification based on a \text{Kronecker} product
  decomposition.
\newblock {\em IEEE/ACM Transactions on Audio, Speech, and Language
  Processing}, 26:1793--1808, 2018.

\bibitem{papadimitriou2021data}
Dimitris Papadimitriou and Swayambhoo Jain.
\newblock Data-driven low-rank neural network compression.
\newblock In {\em IEEE International Conference on Image Processing}, 2021.

\bibitem{peng2024graph}
Chen Peng, Di~Zhang, and Urbashi Mitra.
\newblock Graph identification and upper confidence evaluation for causal
  bandits with linear models.
\newblock In {\em ICASSP 2024-2024 IEEE International Conference on Acoustics,
  Speech and Signal Processing (ICASSP)}, pages 7165--7169. IEEE, 2024.

\bibitem{reinsel2022multivariate}
Gregory~C Reinsel, Raja~P Velu, and Kun Chen.
\newblock {\em Multivariate Reduced-Rank Regression: Theory, Methods and
  Applications}, volume 225.
\newblock Springer Nature, 2022.

\bibitem{saad2003iterative}
Yousef Saad.
\newblock {\em Iterative methods for sparse linear systems}.
\newblock SIAM, 2003.

\bibitem{stewart2001matrix}
Gilbert~W Stewart.
\newblock {\em Matrix Algorithms: Volume \text{II}: Eigensystems}.
\newblock SIAM, 2001.

\bibitem{tai2015convolutional}
Cheng Tai, Tong Xiao, Yi~Zhang, Xiaogang Wang, et~al.
\newblock Convolutional neural networks with low-rank regularization.
\newblock In {\em International Conference on Learning Representations}, 2016.

\bibitem{wang2025svdllm}
Xin Wang, Yu~Zheng, Zhongwei Wan, and Mi~Zhang.
\newblock {SVD}-{LLM}: Truncation-aware singular value decomposition for large
  language model compression.
\newblock In {\em International Conference on Learning Representations}, 2025.

\bibitem{wen2018survey}
Fei Wen, Lei Chu, Peilin Liu, and Robert~C Qiu.
\newblock A survey on nonconvex regularization-based sparse and low-rank
  recovery in signal processing, statistics, and machine learning.
\newblock {\em IEEE Access}, 6:69883--69906, 2018.

\bibitem{xu2019trained}
Yuhui Xu, Yuxi Li, Shuai Zhang, Wei Wen, Botao Wang, Wenrui Dai, Yingyong Qi,
  Yiran Chen, Weiyao Lin, and Hongkai Xiong.
\newblock Trained rank pruning for efficient deep neural networks.
\newblock In {\em Workshop on Energy Efficient Machine Learning and Cognitive
  Computing}, 2019.

\bibitem{yu2017compressing}
Xiyu Yu, Tongliang Liu, Xinchao Wang, and Dacheng Tao.
\newblock On compressing deep models by low rank and sparse decomposition.
\newblock In {\em Conference on Computer Vision and Pattern Recognition}, 2017.

\bibitem{yuan2007dimension}
Ming Yuan, Ali Ekici, Zhaosong Lu, and Renato Monteiro.
\newblock Dimension reduction and coefficient estimation in multivariate linear
  regression.
\newblock {\em Journal of the Royal Statistical Society Series B: Statistical
  Methodology}, 69:329--346, 2007.

\bibitem{yuan2023asvd}
Zhihang Yuan, Yuzhang Shang, Yue Song, Qiang Wu, Yan Yan, and Guangyu Sun.
\newblock Asvd: Activation-aware singular value decomposition for compressing
  large language models.
\newblock {\em ArXiv:2312.05821}, 2023.

\bibitem{zhang2023adaptive}
Lantian Zhang and Lei Guo.
\newblock Adaptive identification with guaranteed performance under saturated
  observation and nonpersistent excitation.
\newblock {\em IEEE Transactions on Automatic Control}, 69:1584--1599, 2023.

\end{thebibliography}

\end{document}